\documentclass[preprint,preprintnumbers,amsmath,amssymb,10pt]{revtex4}
\usepackage[colorlinks,linkcolor=red,citecolor=blue]{hyperref}
\usepackage{epsf,epsfig}
\usepackage{bm}
\usepackage{graphicx}
\usepackage{natbib}
\usepackage{color}
\usepackage{mathrsfs}
\usepackage{subfigure}
\usepackage{multirow}
\usepackage{makecell}
\usepackage{booktabs}
\usepackage{doi}

\newcommand{\commentold}[1]{}
\DeclareMathSymbol{:}{\mathpunct}{operators}{"3A}

\begin{document}


\title{Hamilton-Jacobi analysis of noncanonical inflation in $f(R, T)$ gravity: Constraints from Planck/ACT data, and theoretical bounds}

\author{Z. Ossoulian\footnote{zossoulian@gmail.com}, T. Golanbari\footnote{t.golanbari@uok.ac.ir; t.golanbari@gmail.com}, and Kh. Saaidi\footnote{ksaaidi@uok.ac.ir}}
\affiliation{\small{Department of Physics, University of Kurdistan, P.O. Box 66177-15175, Sanandaj, Iran}}
\date{\today}
\begin{abstract}
	The latest CMB data from ACT DR6, in combination with Planck, DESI, and BICEP/Keck, indicate a slight upward shift in the scalar spectral index. This trend puts several previously favored inflationary models under tension. In this work, we study an inflationary scenario in the framework of $f(R, T)$ gravity, where $R$ is the Ricci scalar and $T$ is the trace of the energy-momentum tensor, with a nonminimal coupling between matter and curvature. The inflaton is assumed to be a noncanonical scalar field with a generalized kinetic energy. To analyze the dynamics of inflation, we employ the Hamilton-Jacobi formalism, where the Hubble parameter is expressed as a function of the scalar field rather than the potential. Within this setup, we examine two functional forms of the Hubble parameter, a power-law and an exponential form, and derive key observables such as the scalar spectral index $n_s$ and the tensor-to-scalar ratio $r$. Comparing the results with ACT DR6, we explore the parameter space of the model. We find that the power-law case fits the data well across a wide range of free parameters, while the exponential case requires a large number of e-folds to be consistent with observations. After inflation, we study reheating, where the dynamics of reheating and inflation are closely linked. Taking into account the overproduction of primordial gravitational waves constrained by the observational bound on $\Delta N_{\text{eff}}$, we obtain a lower limit on the reheating temperature, which is especially restrictive for the stiff equation of state $\omega_{\text{re}}$. This bound implies that the number of e-folds of inflation should generally not exceed $N \lesssim 64(65)$. The resulting energy spectrum of gravitational waves exhibits an enhanced amplitude, thereby bringing it within the observable range of upcoming detectors. We also check the consistency of the model with the Swampland conjectures and the Trans-Planckian Censorship Conjecture (TCC). Our results demonstrate that combining $f(R, T)$ gravity with noncanonical field dynamics provides a rich and testable framework for the early universe. In addition, the Hamilton-Jacobi approach, by avoiding extra approximations, yields a clearer picture of inflation in modified gravity and opens new directions for addressing fundamental problems in high-energy cosmology.

	\vspace{1cm}
\end{abstract}

\maketitle

\section{Introduction}\label{Sec_intro}

General Relativity (GR), proposed by Einstein in 1915 \cite{einstein1915,misner1973gravitation}, has been a remarkably successful theory of gravitation, confirmed by numerous experimental tests across both local and cosmological scales. However, when applied to the evolution of the universe, GR encounters several challenges, including the horizon, flatness, and monopole problems inherent in standard Big Bang cosmology \cite{kolb1990early}. These issues motivated the development of the inflationary paradigm: a brief period of accelerated expansion in the early universe. This idea was first introduced by Starobinsky through quantum corrections to gravity \cite{starobinsky1980new}, and later developed by Guth \cite{Guth:1980zm} as a solution to the flatness and horizon problems, Linde \cite{linde1982new,linde1983chaotic}, and Albrecht and Steinhardt \cite{Albrecht:1982mp}. For comprehensive reviews of inflationary models, see e.g. \cite{lyth1999particle}.

Inflation not only resolves these cosmological problems but also provides a natural mechanism for generating primordial density perturbations that seeded the large-scale structure of the universe. The generation of such perturbations was first proposed in \cite{Mukhanov:1981xt} and has since been extensively studied (for a comprehensive review on cosmological perturbations, refer to \cite{Mukhanov:1992}). This framework has been especially well-developed in the context of single-field slow-roll inflation \cite{Linde:2000kn,Riotto:2002yw,Baumann:2009ds,lyth1999particle}. Over time, the inflationary scenario has gained strong observational support. In particular, the Planck satellite measurements \cite{Planck:2013jfk,Akrami:2018odb,Planck:2023} confirmed the nearly scale-invariant spectrum of scalar perturbations predicted by inflation. More recently, combined analyses from ACT DR6 together with Planck, DESI, BAO, and BICEP/Keck data have tightened constraints on inflationary observables: the scalar spectral index has shifted upward to $n_s = 0.9743 \pm 0.0034$, and the tensor-to-scalar ratio is now bounded by $r \lesssim 0.032$ at 95\% confidence level \cite{ACT:2023dr6,ACT:2025dr6}. These results place increasing tension on standard benchmark models such as Starobinsky and metric Higgs-like inflation.

The new results motivate reconsidering the inflationary scenario within modified gravity frameworks \cite{Chakraborty:2025oyj,Yogesh:2025wak,Gao:2025viy,Pallis:2025nrv}, although certain specific potentials may still remain consistent with the data \cite{Dioguardi:2025vci,Mohammadi:2025gbu,Burikham:2024}.

Beyond the standard models with canonical kinetic terms, several generalized inflationary scenarios have been introduced, including those with noncanonical kinetic terms such as k-inflation  \cite{ArmendarizPicon:1999rj,ArmendarizPicon:2000dh,ArmendarizPicon:2000ah,Mohammadi:2019qeu}, Dirac-Born-Infeld (DBI) inflation \cite{Silverstein:2003hf,Alishahiha:2004eh,Mohammadi:2024bye,Mohammadi:2018zkf,Nazavari:2016yaa}, and approaches based on the Hamilton-Jacobi formalism \cite{Salopek:1990jq,Kinney:1997ne,Sayar:2017pam,Akhtari:2017mxc}. Such extensions provide a broader phenomenological landscape that can better accommodate high-energy physics models and allow more flexibility in fitting cosmological data \cite{Weinberg:2013xma,Renaux-Petel:2015mga,Baumann:2014nda,Mohammadi2021BraneTCC,Mohammadi2020ConstantRoll,Mohammadi2022}. At the same time, the discovery of the late-time acceleration of the universe has presented additional challenges for GR. In the standard cosmological model, this phenomenon is explained by dark energy, typically modeled as a cosmological constant. However, the associated fine-tuning and coincidence problems have motivated interest in alternative approaches, especially those involving modified gravity theories \cite{Clifton:2011jh,Nojiri:2010wj,Capozziello:2011et,Joyce:2014kja,Koyama:2015vza,Ishak:2018his,Mohammadi2022ScalarTensor,Khan:2022odn,Gangopadhyay:2022vgh,Yogesh:2024mpa,Yogesh:2024vcl}.

Among the various modifications, $f(R,T)$ gravity has attracted significant attention. In this theory, the gravitational action is generalized to an arbitrary function of the Ricci scalar $R$ and the trace of the energy-momentum tensor $T$, introducing a coupling between matter and geometry that leads to novel dynamical features \cite{Harko:2011kv}. A wide range of cosmological applications of $f(R,T)$ gravity have been investigated, including its implications for dark energy \cite{Tretyakov:2018,Baffou:2021,Moraes:2017rrv}, dark matter modeling \cite{Zaregonbadi:2016xna}, wormhole solutions \cite{Moraes:2017pao}, gravitational waves \cite{Alves:2016iks}, and inflation \cite{Bhattacharjee:2020jsf,Gamonal:2020itt,Mohammadi:2023kzd}. More recently, both observational and theoretical developments have placed stronger constraints on $f(R,T)$ models \cite{Myrzakulov:2025DE,Tangphati:2024,Bhardwaj:2022,Sharma:2024,Jeakel:2023fRT,Asghari:2025GWfRT,Myrzakulov:2023LambdaT,Bouali:2023obsfRT}. 

In this work, we propose a new class of inflationary models within the framework of $f(R,T)$ gravity, where inflation is driven by a noncanonical scalar field. To analyze the dynamics, the Hamilton-Jacobi formalism in employed, in which the Hubble parameter is expressed as a function of the scalar field, allowing for a first-order treatment of inflationary evolution \cite{Salopek:1990jq,Kinney:1997ne}. We focus on two specific forms of the Hubble parameter, namely the power-law and exponential types, and derive the corresponding inflationary observables, including the scalar spectral index $n_s$ and the tensor-to-scalar ratio $r$. By comparing the model predictions with ACT DR6 results, the parameter space of the model is constrained. A key feature of the framework is the role of the coupling constant $\lambda$, which affects the sound speed. Increasing $\lambda$ reduces the sound speed, and since $r$ is proportional to it, this naturally suppresses the tensor-to-scalar ratio. As a consequence, potentials that are excluded in standard inflationary scenarios, such as the power-law potential, can become consistent with observations within $f(R,T)$ gravity. 

After inflation, the universe enters a supercooled state, nearly empty of particles. A reheating phase is therefore required to fill the universe with standard particles, ensuring a smooth transition to the standard radiation-dominated epoch \cite{Kofman:1994rk,Kofman:1997yn,Allahverdi:2010xz,Cook:2015vqa,Dai:2014jja,Munoz:2014eqa}. The reheating process is described by the temperature $T_{\text{re}}$, the reheating e-folds $N_{\text{re}}$, and the effective equation-of-state parameter $\omega_{\text{re}}$ \cite{Martin:2014nya,Rehagen:2015zma}. In general, there is a wide acceptable range for the reheating temperature, so that it should stand between the upper bound at the GUT scale, $T_{GUT} \sim 10^{16}\,\mathrm{GeV}$, and the lower bound from BBN, $T_{BBN} \sim 4\,\mathrm{MeV}$ \cite{Fields:2019pfx,Cyburt:2015mya}. In addition to these bounds, primordial gravitational waves (PGWs), generated as tensor perturbations during inflation, provide additional constraints. PGWs freeze after horizon crossing but re-enter after inflation and evolve. The earlier modes that enter the reheating phase can significantly affect PGWs, especially for modes re-entering the horizon, leading to an enhancement in the high-frequency spectrum and a blue tilt \cite{Nakayama:2008wy,Boyle:2005se,Kuroyanagi:2009br}. Consequently, the total PGW energy density increases, modifying the effective number of relativistic species, $\Delta N_{\text{eff}}$. Observational bounds from BBN and CMB impose strong constraints on $\Delta N_{\text{eff}}$, providing a powerful handle on the dynamics of reheating.

Finally, we analyze the theoretical viability of our model in light of recent developments in quantum gravity, including the Swampland Conjectures \cite{PhysRevD.98.086004,Garg:2018reu,Ooguri:2018Refined,Palti:2019Review} and the TCC \cite{Bedroya:2019TCC,Bedroya:2020TCCInflation}. Our results show that the proposed model is both observationally consistent and theoretically viable, making it a promising candidate for describing early-universe inflation within a modified gravity framework.

In summary, while earlier studies of inflation in $f(R,T)$ gravity have primarily considered canonical scalar fields or employed restricted functional forms, our work develops a broader framework by incorporating a non-canonical scalar field and analyzing it within the Hamilton-Jacobi formalism. To our knowledge, this is the first systematic Hamilton-Jacobi study of noncanonical inflation in $f(R,T)$ gravity that directly confronts ACT DR6 observational data and is tested against fundamental theoretical criteria, such as the Swampland conjectures and the TCC. This simultaneous confrontation with both observations and theory highlights the novelty of our approach and its potential to provide new insights into early-universe cosmology. The structure of the paper is as follows. Section~\ref{fRT_gravity} reviews the general framework of $f(R,T)$ gravity. Section~\ref{NC_inflation} introduces the noncanonical inflationary setup together with the Hamilton-Jacobi formalism and the associated perturbation parameters. Section~\ref{reheating} analyzes the reheating dynamics, including the impact of primordial gravitational waves (PGWs) and the constraints from the effective number of relativistic species. In Section~\ref{examples}, we investigate two representative cases of the Hubble parameter and confront their predictions with observational data. Section~\ref{SC-TCC} examines the implications of the Swampland conjectures and the Trans-Planckian Censorship Conjecture (TCC). Finally, Section~\ref{conclusion} summarizes our main findings.

\section{Basic equations in $f(R,T)$ gravity}\label{fRT_gravity} 

We consider the general action of $f(R,T)$ gravity, introduced as an extension of general relativity to include explicit matter–curvature couplings \cite{Harko:2011kv}:
\begin{equation}
S = \frac{1}{2 \kappa^2} \int d^4x \sqrt{-g} \left[ f(R, T) + \mathcal{L}_m \right],
\end{equation}
where $\kappa^2 = 8\pi G$ and $G$ is Newton’s gravitational constant. The function $f(R,T)$ depends on both the Ricci scalar $R$ and the trace of the energy–momentum tensor $T$, while $\mathcal{L}_m$ denotes the matter Lagrangian density. 

Variation of this action with respect to the metric yields the field equations \cite{Houndjo:2011tu,Tretyakov:2018,Baffou:2021,Shabani:2013FRT,Zubair:2015FRTFieldEq,Nojiri:2017ncd}:
\begin{equation}\label{field_equation}
\Xi_{\mu\nu} f_{,R} + f_{,R} R_{\mu\nu} - \frac{1}{2} g_{\mu\nu} f 
= \kappa^2 T_{\mu\nu} - f_{,T} \left( T_{\mu\nu} + \Theta_{\mu\nu} \right),
\end{equation}
where the operators are defined as
\begin{align}
\Xi_{\mu\nu} &= g_{\mu\nu} \square - \nabla_\mu \nabla_\nu, \\
\Theta_{\mu\nu} &= g^{\alpha\beta} \frac{\delta T_{\alpha\beta}}{\delta g^{\mu\nu}}.
\end{align}

Assuming that the matter sector can be modeled as a perfect fluid, the energy–momentum tensor takes the form
\begin{equation}
T_{\mu\nu} = (\rho + p) u_\mu u_\nu - p g_{\mu\nu},
\end{equation}
where $\rho$ and $p$ denote the energy density and pressure, respectively. This setup allows the dynamics to be specialized to a cosmological background.

For analytical simplicity, we adopt the widely studied linear form of the gravitational function \cite{Moraes_2016,Moraes:2015uxq,Carvalho:2017pgk,Moraes:2017rrv,Moraes:2017mir,Moraes:2016akv}:
\begin{equation}
f(R,T) = R + \eta T,
\end{equation}
where $\eta$ is a constant parameter. Introducing a dimensionless coupling $\lambda$ through $\eta = \lambda \kappa^2$, this parameter quantifies the departure from general relativity and determines the strength of the matter–geometry interaction.

In a spatially flat FLRW background, the modified Friedmann equations are obtained as
\begin{align}
H^2 &= \frac{\kappa^2}{3} \left[ \left( \frac{3}{2} \lambda + 1 \right) \rho - \frac{\lambda}{2} p \right], \label{friedmann1} \\
-3H^2 - 2\dot{H} &= \kappa^2 \left[ -\frac{\lambda}{2} \rho + \left( \frac{3}{2} \lambda + 1 \right) p \right], \label{friedmann2}
\end{align}
where $H = \dot{a}/a$ is the Hubble parameter. These equations clearly show that the presence of $\lambda$ modifies the effective contributions of energy density and pressure, thereby leading to deviations from the standard GR background evolution.

Combining Eqs.~\eqref{friedmann1} and \eqref{friedmann2} gives the evolution equation for the Hubble parameter:
\begin{equation}\label{dHt}
-2 \dot{H} = \kappa^2 (1+\lambda)(\rho + p).
\end{equation}

Differentiating Eq.~\eqref{friedmann1} and substituting Eq.~\eqref{dHt}, one obtains the modified conservation law:
\begin{equation}\label{modified_conservation}
\left( \frac{3\lambda}{2} + 1 \right) \dot{\rho} - \frac{\lambda}{2} \dot{p} + 3H(1 + \lambda)(\rho + p) = 0.
\end{equation}
This result demonstrates that, owing to the explicit matter–geometry coupling, the energy–momentum tensor is not separately conserved in $f(R,T)$ gravity. 

Finally, in the limit $\lambda \to 0$, all equations smoothly recover the corresponding results of general relativity, as expected.

\section{Noncanonical inflationary dynamics}\label{NC_inflation} 

We next consider a noncanonical scalar field whose dynamics is governed by the generalized Lagrangian
\begin{equation}\label{action}
\mathcal{L}(\phi, X) = X \left( \frac{X}{M^4} \right)^{\alpha -1} - V(\phi),
\end{equation}
where $X=\dot{\phi}^2/2$ denotes the kinetic term of the scalar field and $M$ denotes the mass scale in the noncanonical kinetic term, while the inflationary energy scale is identified with $\rho^{1/4}$ at horizon crossing. This form belongs to a broad class of $k$-essence models, in which the parameter $\alpha$ controls the kinetic contribution. The canonical case is recovered for $\alpha=1$, while values $\alpha>1$ introduce departures from standard dynamics. 

From Eq.~\eqref{action}, the scalar field energy density and pressure are obtained as
\begin{align}
\rho &= (2\alpha - 1) X \left( \frac{X}{M^4} \right)^{\alpha -1} + V(\phi), \nonumber \\
p &= X \left( \frac{X}{M^4} \right)^{\alpha -1} - V(\phi). \label{energy_pressure}
\end{align}

Substituting these expressions into the modified Friedmann equations \eqref{friedmann1} and \eqref{dHt}, one finds
\begin{align}
H^2 &= \frac{1}{3M_p^2} \Bigg[ \Bigg( 2\alpha \Big( \tfrac{3\lambda}{2} + 1 \Big) - (1 + 2\lambda) \Bigg) 
\left( \frac{X}{M^4} \right)^{\alpha -1} X 
+ (1 + 2\lambda)\, V(\phi) \Bigg], \nonumber \\
\dot{H} &= - \frac{1}{2 M_p^2} (1 + \lambda)(2\alpha) \left( \frac{X}{M^4} \right)^{\alpha -1} X. \label{dHt_nc}
\end{align}
These relations show that the coupling parameter $\lambda$ modifies the relative weights of the kinetic and potential contributions to the expansion rate, leading to deviations from both general relativity and standard $k$-inflation scenarios.

The scalar field equation of motion, derived from the modified conservation law \eqref{modified_conservation}, takes the form
\begin{equation}\label{field_eom}
\Bigg( 2\alpha \Big( \tfrac{3\lambda}{2} + 1 \Big) - (1 + 2\lambda) \Bigg)\ddot{\phi}
+ 3 H (1 + 2\lambda) \dot{\phi}
+ \frac{1}{\alpha} \left( \frac{M^4}{X} \right)^{\alpha - 1} V'(\phi) = 0,
\end{equation}
where the prime denotes differentiation with respect to $\phi$. The coefficients of both the acceleration term $\ddot{\phi}$ and the Hubble friction term $3H\dot{\phi}$ depend explicitly on $\alpha$ and $\lambda$, showing that the effective inertia and damping of the scalar field are modified by the noncanonical kinetic structure and the matter–geometry coupling. In the special case $\alpha=1$ and $\lambda=0$, Eq.~\eqref{field_eom} reduces to the standard Klein–Gordon equation. 

\subsection{Hamilton-Jacobi formalism}

In the Hamilton-Jacobi formalism, instead of treating the potential as a function of the scalar field, the Hubble parameter is directly expressed as a function of $\phi$, namely $H=H(\phi)$. This formulation provides a convenient framework to study inflationary dynamics without assuming an explicit form of $V(\phi)$. 
Since $\dot{H} = \dot{\phi}\, H'(\phi)$, substitution into Eq.~\eqref{dHt_nc} gives
\begin{equation}\label{Hprime}
H'(\phi) = - \frac{\alpha (1 + \lambda)}{M_p^2 \, \xi} \; \frac{X^\alpha}{\dot{\phi}},
\end{equation}
where $\xi = M^{4(\alpha - 1)}$ is constant. The condition $\epsilon_1>0$ implies $\alpha(1+\lambda)>0$, hence $H'$ and $\dot{\phi}$ must carry opposite signs. We adopt $\dot{\phi}<0$, corresponding to the field rolling down its effective potential. Equation~\eqref{Hprime} thus connects the slope of the Hubble parameter to the kinetic structure of the scalar field, with both $\alpha$ and $\lambda$ playing central roles. 

By eliminating $X$ from Eq.~\eqref{dHt_nc} and substituting into the Friedmann equation, the potential can be expressed in terms of the Hubble parameter as
\begin{equation}\label{HJequation}
V(\phi) = \frac{3M_p^2}{(1+2\lambda)} \, H^2 
\left[ 1 - \frac{2\alpha \big( \tfrac{3 \lambda}{2} + 1 \big) - (1 + 2\lambda)}{3\alpha(1+\lambda)} \, \epsilon_1 \right],
\end{equation}
This relation represents the generalized Hamilton–Jacobi potential in $f(R,T)$ gravity with a noncanonical scalar field. 
In the canonical limit $(\alpha=1,\lambda=0)$, this reduces to $V=M_p^2H^2(3-\epsilon_1)$, in agreement with standard general relativity.
The first slow-roll parameter, which characterizes the variation of the Hubble parameter during inflation, is given by
\begin{equation}\label{first_srp}
\epsilon_1 = -\frac{\dot{H}}{H^2} 
= \left( \frac{2^\alpha M_p^2 \xi}{\alpha (1+\lambda)} \right)^{\!\tfrac{1}{2\alpha - 1}} 
\frac{H'(\phi)^{\tfrac{2\alpha}{2\alpha - 1}}}{H^2}.
\end{equation}
This parameter depends nontrivially on both $\alpha$ and $\lambda$, while reducing to the canonical expression when $\alpha=1$ and $\lambda=0$. 
A second slow-roll parameter is defined to track the evolution of $\epsilon_1$,
\begin{equation}\label{second_srp}
\epsilon_2 = \frac{\dot{\epsilon}_1}{H \epsilon_1} = 2 \epsilon_1 - 2\alpha \, \eta_H,
\end{equation}
with the auxiliary quantity $\eta_H$ defined as
\begin{equation}\label{eta_srp}
\eta_H = \frac{1}{2\alpha - 1} 
\left( \frac{2^\alpha M_p^2 \xi}{\alpha(1+\lambda)} \right)^{\!\tfrac{1}{2\alpha - 1}} 
\frac{H'(\phi)^{\tfrac{2-2\alpha}{2\alpha - 1}} H''(\phi)}{H}.
\end{equation}
This parameter encodes the curvature of the Hubble parameter and quantifies deviations from exact de Sitter expansion. 
Finally, the number of $e$-folds between horizon exit ($\phi_\star$) and the end of inflation ($\phi_e$) is
\begin{equation}\label{efolds}
N = \int_{\phi_\star}^{\phi_e} \frac{H(\phi)}{\dot{\phi}} \, d\phi.
\end{equation}
Using the expression for $\dot{\phi}$ from Eq.~\eqref{Hprime}, one obtains
\begin{equation}
\dot{\phi} = \left[ - \frac{M_p^2 \xi \, 2^\alpha}{\alpha(1 + \lambda)} H'(\phi) \right]^{\!\tfrac{1}{2\alpha - 1}},
\end{equation}
which leads to the final integral expression
\begin{equation}
N = \int_{\phi_\star}^{\phi_e} H(\phi) 
\left[ - \frac{M_p^2 \xi \, 2^\alpha}{\alpha(1 + \lambda)} H'(\phi) \right]^{-\tfrac{1}{2\alpha - 1}} 
\, d\phi.
\end{equation}
This relation will be evaluated explicitly for specific choices of $H(\phi)$ in Sec.~\ref{examples}.

\subsection{Perturbations}

To assess the observational viability of the inflationary model, it is necessary to study the perturbations generated during the inflationary epoch. In this subsection, we introduce the key quantities related to scalar and tensor perturbations, which provide the basis for comparison with observational constraints in Sec.~\ref{examples}. A central observable is the amplitude of scalar perturbations, which for the present model reads
\begin{equation}\label{Ps}
\mathcal{P}_s = \frac{1}{4\pi^2} \; \frac{H^4}{c_s \big( \rho_{\text{eff}} + p_{\text{eff}} \big)},
\end{equation}
where $\rho_{\text{eff}}$ and $p_{\text{eff}}$ are derived from the modified Friedmann equations \eqref{friedmann1} and \eqref{friedmann2} as
\begin{align*}
\rho_{\text{eff}} &= \left( \frac{3\lambda }{2}+ 1 \right) \rho - \frac{\lambda}{2} \, p, \\
p_{\text{eff}} &= - \frac{\lambda}{2} \, \rho + \left( \frac{3\lambda }{2}+ 1 \right) p,
\end{align*}
with $\rho$ and $p$ denoting the energy density and pressure of the noncanonical scalar field given in Eq.~\eqref{energy_pressure}. Because of the Lagrangian form in Eq.~\eqref{action}, the combination $T+\mathcal{L}_m$ can be expressed in terms of $X=\dot{\phi}^2/2$ and $V(\phi)$. This places the model within the class of $k$-essence theories, whose perturbative analysis was originally developed in \cite{Garriga:1999vw}. 
The effective sound speed $c_s$ of scalar perturbations is defined as
\begin{equation}\label{sound_speed}
c_s^2 = \frac{\dot{p}_{\text{eff}}}{\dot{\rho}_{\text{eff}}} =
\frac{(1 + 2\lambda) - \alpha \lambda}{2\alpha \left( \tfrac{3 \lambda}{2} + 1 \right) - \left( 1 + 2\lambda \right)},
\end{equation}
In this model the sound speed is constant, simplifying the perturbation analysis compared to general $k$-essence. In the canonical limit ($\alpha=1,\lambda=0$) one recovers $c_s^2=1$, while more generally $\alpha=1$ yields $c_s^2=1$ for any $\lambda$, corresponding to a canonical scalar in modified gravity. At $\lambda=-1$ the expression is formally $0/0$, but the limit $\lambda\to -1$ still gives $c_s^2=1$.

\begin{figure}[h]
\centering 
\includegraphics[width = 0.6 \linewidth]{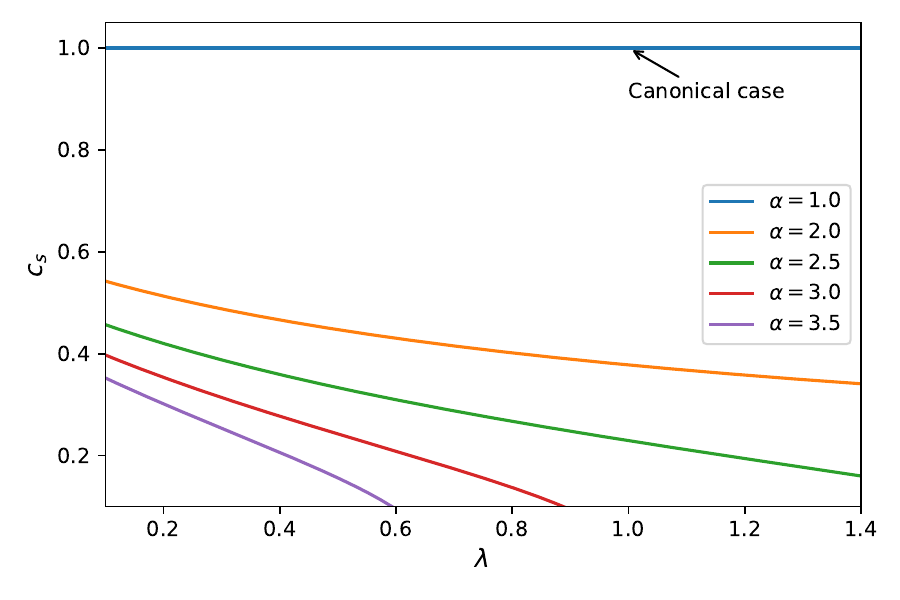}
\caption{The sound speed $c_s$ as a function of $\lambda$ for different values of $\alpha$. For $\alpha=1$ (canonical case), one has $c_s^2=1$ independently of $\lambda$, while for larger $\alpha$ the sound speed decreases with increasing $\lambda$.}
\label{soundspeed}
\end{figure}

As shown in Fig.~\ref{soundspeed}, the sound speed decreases with increasing values of both $\alpha$ and $\lambda$. Since the tensor-to-scalar ratio depends linearly on $c_s$, this mechanism allows the model to predict smaller values of $r$ (see Eq.~\eqref{r}). In particular, this feature enables power-law potentials to remain consistent with current observational limits, including those reported by the ACT collaboration (see Fig.~10 of \cite{ACT:2025dr6}).

The scalar spectral index, which characterizes the scale dependence of the scalar power spectrum, is given by
\begin{equation}\label{ns}
n_s = 1 - (2\epsilon_1 + \epsilon_2),
\end{equation}
where $\epsilon_1$ and $\epsilon_2$ are the slow-roll parameters defined in the Hamilton–Jacobi formalism of the previous subsection. Another key observable is the tensor-to-scalar ratio,
\begin{equation}\label{r}
r = 16 \; c_s \; \epsilon_1.
\end{equation}
The expressions for $n_s$ and $r$ constitute the main theoretical predictions of the model. In Sec.~\ref{examples}, these predictions will be confronted with the most recent cosmological data, in particular those from Planck 2018 and ACT DR6.

\section{Reheating}\label{reheating}

During the rapid accelerated expansion of inflation, the universe enters a super-cooled state with almost no standard particles. To recover the radiation-dominated epoch, a mechanism is required to repopulate the universe with particles and restore a hot thermal state. This transition, referred to as reheating, transfers the inflaton energy into particle degrees of freedom, which subsequently thermalize and establish the hot Big Bang phase 
\cite{Albrecht:1982mp,Traschen:1990sw,Shtanov:1994ce,Kofman:1997yn,Bassett:2005xm,Allahverdi:2010xz,Rehagen:2015zma,Cook:2015vqa,Dai:2014jja,Munoz:2014eqa,Ueno:2016dim,Figueroa:2021yhd,Yadav:2024MHI,German:2024AlphaReheating,Asfour:2024HiggsHybrid,deHaro:2024GravReheat,German:2025AlphaDynamics,Liu:2025HiggsACT}. 

The dynamics of reheating are conveniently parametrized by three quantities \cite{Martin:2014nya,Rehagen:2015zma}:  
(i) the number of reheating $e$-folds, $N_{\text{re}}$,  
(ii) the reheating temperature, $T_{\text{re}}$, and  
(iii) the effective equation of state $\omega_{\text{re}}$ during the reheating stage.  
Assuming a constant $\omega_{\text{re}}$ and expressing the final energy density as $\rho_{re} = \tfrac{\pi^2}{30} g_{\star re} T_{\text{re}}^4$, one obtains
\begin{equation}\label{Nre}
N_{\text{re}} = -\frac{1}{3(1+\omega_{\text{re}})}\ln\!\left(\frac{\pi^2 g_{\star re}}{90 M_p^2 H_e^2}\right) - \frac{4}{3(1+\omega_{\text{re}})}\ln\!\left(T_{\text{re}}\right),
\end{equation}
which links the reheating duration to $T_{\text{re}}$, $\omega_{\text{re}}$, and the inflationary scale $H_e$. 
A second relation follows from entropy conservation:
\begin{equation}\label{Tre1}
T_{\text{re}} = \left( \frac{43}{11 g_{\star s, re}} \right)^{1/3} 
\frac{H_k \, T_0}{k_\star} \; e^{-(N_k + N_{\text{re}})},
\end{equation}
where $g_{\star s,re}$ is the effective number of relativistic species contributing to entropy, $N_k$ denotes the number of $e$-folds between horizon exit and the end of inflation, $H_k$ is the Hubble parameter at horizon crossing, and $T_0=2.735\,\mathrm{K}=2.35\times10^{-4}\,\mathrm{eV}$ is the current CMB temperature (with $a_0=1$). This expression shows explicitly how $T_{\text{re}}$ is controlled by both the inflationary history and the subsequent expansion.  
Combining Eqs.~\eqref{Nre} and \eqref{Tre1} gives a more practical form \cite{Cook:2015vqa}:
\begin{eqnarray}\label{Tre2}
T_{\text{re}} &=& \Bigg[ \left( \frac{11 g_{\star s, re}}{43} \right)^{1/3} 
\left( \frac{90 M_p^2 H_e^2}{\pi^2 g_{\star re}} \right)^{\tfrac{1}{3(1+\omega_{\text{re}})}} 
\frac{k_\star}{H_k \, T_0} \; e^{N_k} \Bigg]^{\tfrac{3(1+\omega_{\text{re}})}{1-3\omega_{\text{re}}}},
\end{eqnarray}
which allows $T_{\text{re}}$ to be directly estimated once the inflationary parameters and $\omega_{\text{re}}$ are specified.
The reheating temperature is expected within a broad but constrained interval. The lower limit is fixed by successful BBN, requiring $T_{BBN} \simeq 4\,\mathrm{MeV}$ \cite{Kawasaki:2005nd,Kawasaki:2000en,Dai:2014jja}, while the upper limit is often associated with the GUT scale, $T_{GUT}\simeq 10^{16}\,\mathrm{GeV}$. Beyond these bounds, primordial gravitational waves (PGWs) generated during inflation can impose stronger constraints \cite{Boyle:2005se,Watanabe:2006qe,Saikawa:2018rcs,Caprini:2018mtu,Bernal:2020ywq,PhysRevD.101.043528}. PGWs re-enter the horizon during reheating, contributing to the radiation energy density. For stiff reheating equations of state ($\omega_{\text{re}}>1/3$), this contribution becomes enhanced and may shift the effective number of relativistic species, $\Delta N_{\rm eff}$. As a result, PGWs can place stringent bounds on both the temperature and duration of reheating. Future gravitational-wave observations are expected to probe these effects with high precision 
\cite{Bernal:2025JHEP_ThermalGW,Haque:2025HiggsStarobinsky,Kanemura:2024GWdecay,Barman:2024GWbrem,Fan:2024PTAreheating}, providing a powerful window into the microphysics of the reheating era. 

\subsection{PGW and the number of relativistic species}

Inflation not only solves the classical problems of the standard Big Bang model but also provides a natural mechanism for generating primordial perturbations. While scalar fluctuations give rise to the large-scale structure of the universe, inflation also predicts a stochastic background of primordial gravitational waves (PGWs). These tensor perturbations can leave an imprint on the effective number of relativistic species, $\Delta N_{\rm eff}$, especially if the post-inflationary universe undergoes a stiff expansion phase with $\omega_{\text{re}} > 1/3$. Since $\Delta N_{\rm eff}$ is tightly constrained by BBN and CMB observations, the evolution of PGWs during reheating offers a sensitive probe of the inflationary model. 

Recent results from \textit{Planck} combined with ACT data impose a stringent bound of $\Delta N_{\rm eff} \leq 0.17$, which translates into an upper limit on the present PGW energy density. Following \cite{Haque:2021dha,Chakraborty:2023ocr,Maity:2024odg}, this constraint can be written as
\begin{equation}
\int_{k_{\text{re}}}^{k_{\text{e}}} \frac{dk}{k} \, \Omega_{\mathrm{GW}}^{(0)}(k) h^2 
\;\leq\; \frac{7}{8} \left( \frac{4}{11} \right)^{4/3} \Omega_{\gamma}^{(0)} h^2 \, \Delta N_{\mathrm{eff}},
\label{pgw1}
\end{equation}
where $\Omega_{\gamma}^{(0)} h^2 \approx 2.47 \times 10^{-5}$ is the present photon energy density, and $k_e$ and $k_{\rm re}$ are the comoving scales associated with the end of inflation and the end of reheating, respectively. Equation \eqref{pgw1} therefore links the present PGW spectrum to observational bounds on extra relativistic degrees of freedom. 
For modes that re-enter the horizon during reheating ($k>k_{\rm re}$), the PGW energy density is amplified, and the effect becomes stronger for $\omega_{\text{re}} > 1/3$. A simplified expression valid during the radiation-dominated era is
\begin{equation}
\Omega_R^{(0)} h^2 \, \frac{H_{\text{e}}^2 \; \mu(\omega_{\text{re}})}{12 \pi M_P^2}  
\frac{(1 + 3 \omega_{\text{re}})^2}{3 \omega_{\text{re}} - 1} 
\left( \frac{k_{\text{e}}}{k_{\text{re}}} \right)^{\frac{6 \omega_{\text{re}} - 2}{1 + 3 \omega_{\text{re}}}}  
\;\leq\; 5.61 \times 10^{-6} \, \Delta N_{\rm eff},
\end{equation}
where $\Omega_R^{(0)}$ is the present radiation density parameter and
\begin{equation}
\mu(\omega_{\text{re}}) = \big( 1 + 3\omega_{\text{re}} \big)^{\tfrac{4}{1+3\omega_{\text{re}}}} \; 
\Gamma^2\!\left( \frac{5+3\omega_{\text{re}}}{2 + 6\omega_{\text{re}}} \right)
\end{equation}
is an order-one factor encoding the dependence on the reheating equation of state. This relation makes it clear that stiff reheating scenarios enhance the PGW background and therefore tighten the bounds on the reheating history. 

Using the relations for $k_e$ and $k_{\rm re}$, one obtains a lower bound on the reheating temperature \cite{Haque:2025uri,Haque:2025uis,Mohammadi:2025PowerLawPlateau,Haque:2021dha}:
\begin{align}
T_{\text{re}} 
\geq 
\left( \frac{90 \, H_{\text{e}}^2 M_P^2}{\pi^2 g_{\star re}}\right)^{\frac{1}{4}} 
\Bigg[ 
\frac{\Omega_R^{(0)} h^2}{5.61 \times 10^{-6} \, \Delta N_{\text{eff}}}
\frac{H_{\text{e}}^2 \; \mu(\omega_{\text{re}})}{12 \pi M_P^2} 
\frac{(1 + 3 \omega_{\text{re}})^2}{3 \omega_{\text{re}} - 1} 
&\Bigg]^{\frac{3(1 + \omega_{\text{re}})}{4 (3 \omega_{\text{re}} - 1)}} 
\nonumber\\
&\equiv T_{\text{re}}^{\rm GW}.
\end{align}
Equations \eqref{pgw1} and its simplified form thus demonstrate how PGWs impose additional lower bounds on the reheating temperature. These bounds complement the results of Sec.~\ref{reheating}, providing a more complete picture of the post-inflationary universe. 

\clearpage
\section{Representative examples}\label{examples} 

To illustrate the implications of the formalism developed in the previous sections, we now consider two representative choices for the Hubble parameter $H(\phi)$.  
The first is a power-law form, $H(\phi)=H_0 \phi^n$, which has long been employed in chaotic inflation scenarios \cite{linde1983chaotic,Lucchin:1985PowerLaw,Unnikrishnan:2012zu}. Such functional forms naturally arise from simple polynomial potentials in effective field theory and have served as benchmark models in both canonical and noncanonical inflationary studies. Within the Hamilton-Jacobi framework, the power-law case provides a clear setting in which slow-roll parameters and perturbative quantities can be evaluated analytically.  

The second example is an exponential form, $H(\phi)=H_0 e^{\gamma \phi}$, often motivated by higher-dimensional and string-inspired theories \cite{Copeland:1998Exp,Copeland:2006wr,Lucchin:1985PowerLaw,Tsujikawa:2013Quintessence}. Exponential functions frequently emerge from the dynamics of moduli or dilaton fields and are known to produce scaling or attractor solutions. These features make them particularly relevant for connecting early-universe inflation to possible late-time accelerated expansion.  

Together, these two cases provide concrete realizations of the general framework developed above, allowing us to explore the inflationary dynamics in detail and to confront the theoretical predictions with current observational constraints. In our work, these realizations are analyzed within the $f(R,T)$ framework with a noncanonical scalar field, thereby extending the standard Hamilton-Jacobi approach to a broader gravitational context.  

\subsection{Case I: Power-law inflation }

As a first concrete realization, we consider the Hubble parameter to follow a power-law dependence on the scalar field,
\begin{equation}
H(\phi) = H_0 \phi^n,
\end{equation}
where $H_0$ and $n$ are constant parameters. Within the Hamilton-Jacobi formalism, the slow-roll parameters take the form
\begin{eqnarray}
\epsilon_1 &=& \left( \frac{2^\alpha M_p^2 \xi}{\alpha (1+\lambda)} \, n^{2\alpha} H_0^{2-2\alpha} \right)^{\tfrac{1}{2\alpha - 1}} 
\, \phi^{\tfrac{2n(1-\alpha) - 2\alpha}{2\alpha - 1}}, \\
\eta_H &=& \frac{n(n-1)}{2\alpha - 1} 
\left( \frac{2^\alpha M_p^2 \xi}{\alpha (1+\lambda)} \, n^{2-2\alpha} H_0^{2-2\alpha} \right)^{\tfrac{1}{2\alpha - 1}} 
\, \phi^{\tfrac{2n(1-\alpha) - 2\alpha}{2\alpha - 1}}.
\end{eqnarray}
The end of inflation is defined by $\epsilon_1=1$, which yields
\begin{equation}\label{phi_end}
\phi_e^{\tfrac{2n(1-\alpha) - 2\alpha}{2\alpha - 1}} 
= \left( \frac{2^\alpha M_p^2 \xi}{\alpha (1+\lambda)} \, n^{2\alpha} H_0^{2-2\alpha} \right)^{-\tfrac{1}{2\alpha - 1}}.
\end{equation}
At horizon crossing, the scalar field value is related to the number of $e$-folds $N$ as
\begin{equation}\label{phi_star}
\phi_\star^{\tfrac{2n(1-\alpha) - 2\alpha}{2\alpha - 1}} 
= \left[ \left( \frac{2^\alpha M_p^2 \xi}{\alpha (1+\lambda)} \, n^{2\alpha} H_0^{2-2\alpha} \right)^{\tfrac{1}{2\alpha - 1}} 
\left( 1 - \frac{2n(1-\alpha) - 2\alpha}{n(2\alpha - 1)} N \right) \right]^{-1}.
\end{equation}
The corresponding slow-roll parameters at horizon exit are then
\begin{eqnarray}
\epsilon_1^\star &=& \left( 1 - \frac{2n(1-\alpha) - 2\alpha}{n(2\alpha - 1)} N \right)^{-1}, \\
\eta_H^\star &=& \frac{n - 1}{n(2\alpha - 1)} \, \epsilon_1^\star,
\end{eqnarray}
which directly determine the scalar spectral index $n_s$ and the tensor-to-scalar ratio $r$ via Eqs.~\eqref{ns} and \eqref{r}.
\begin{figure}[h]
\centering
\begin{minipage}{0.48\textwidth}
	\centering
	\includegraphics[width=\linewidth]{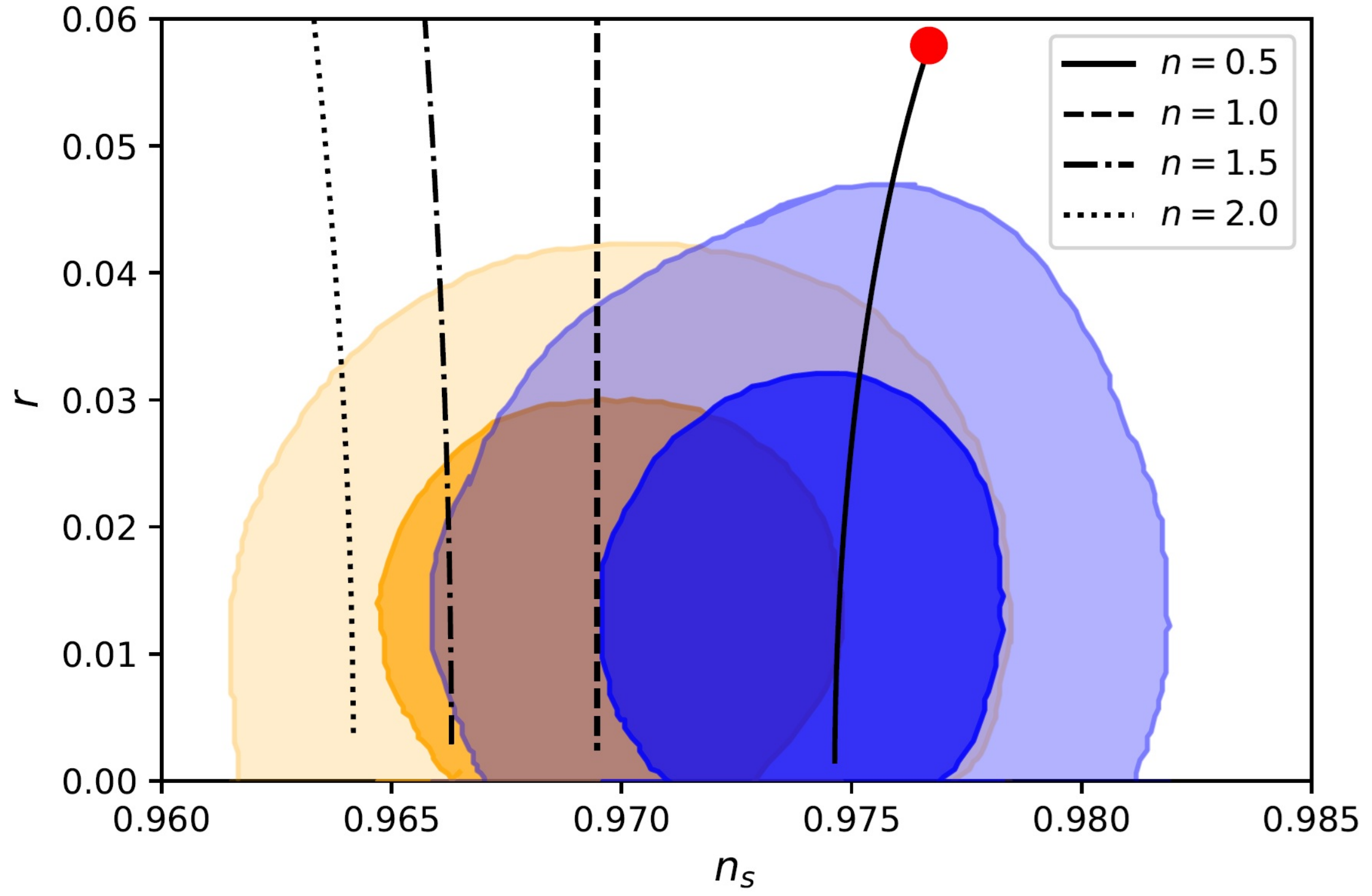}
	(a)\label{pl_rns_vs_a}
\end{minipage}
\hfill
\begin{minipage}{0.48\textwidth}
	\centering
	\includegraphics[width=\linewidth]{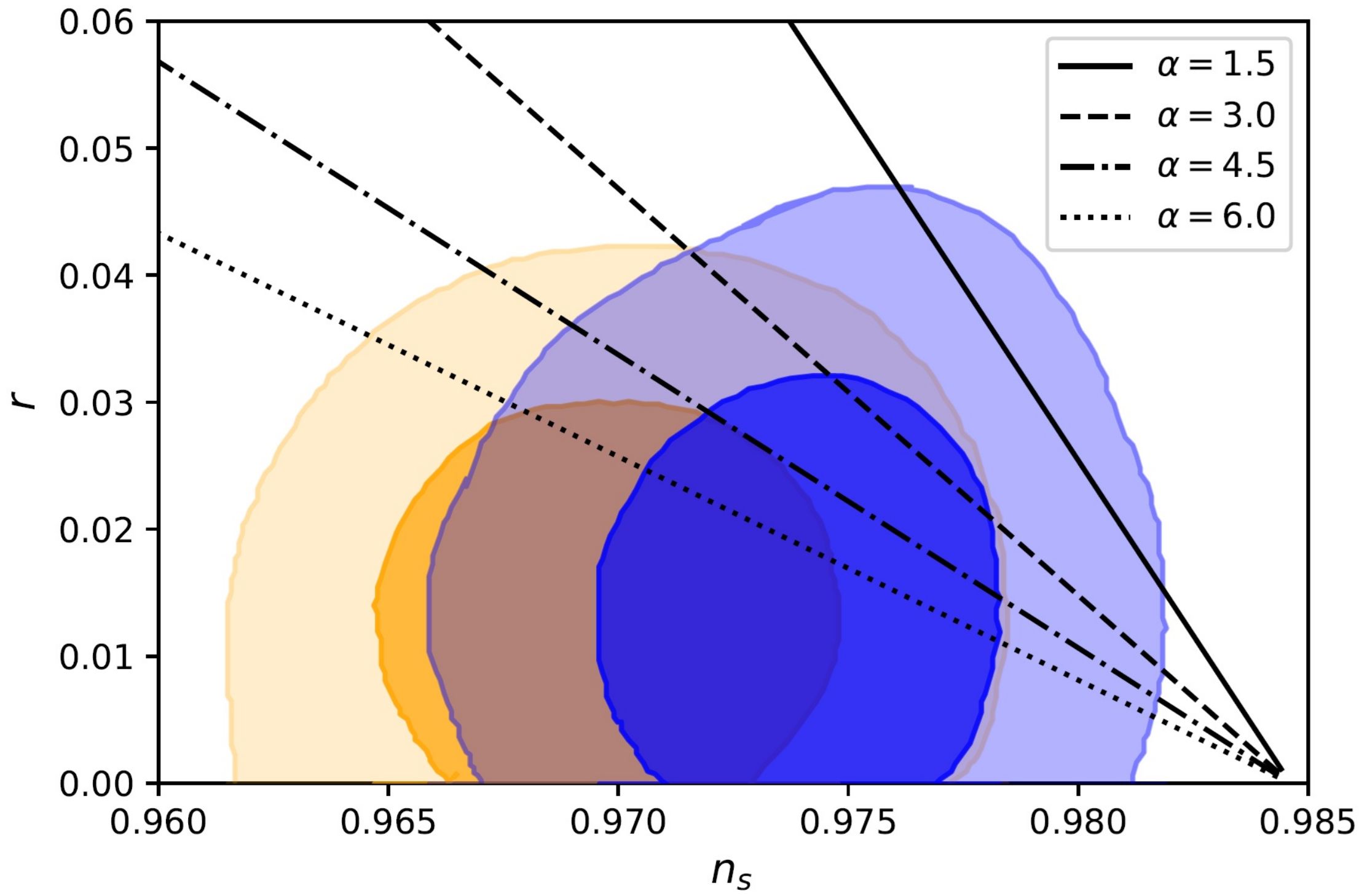}
	(b)\label{pl_rns_vs_n}
\end{minipage}
\caption{Predictions of the power-law case in the $r$--$n_s$ plane: 
	(a) trajectories as a function of $\alpha$ for several $n$, and 
	(b) trajectories as a function of $n$ for several $\alpha$. 
	In both panels we take $N=65$ and $\lambda=0.1$. 
	Shaded regions: orange (Planck 2018, $68\%/95\%$ CL) and blue (ACT DR6, 2025, $68\%/95\%$ CL).}
	\label{pl_rns_vs}
	\end{figure}
	
\begin{figure}
\centering
\includegraphics[width=0.5\linewidth]{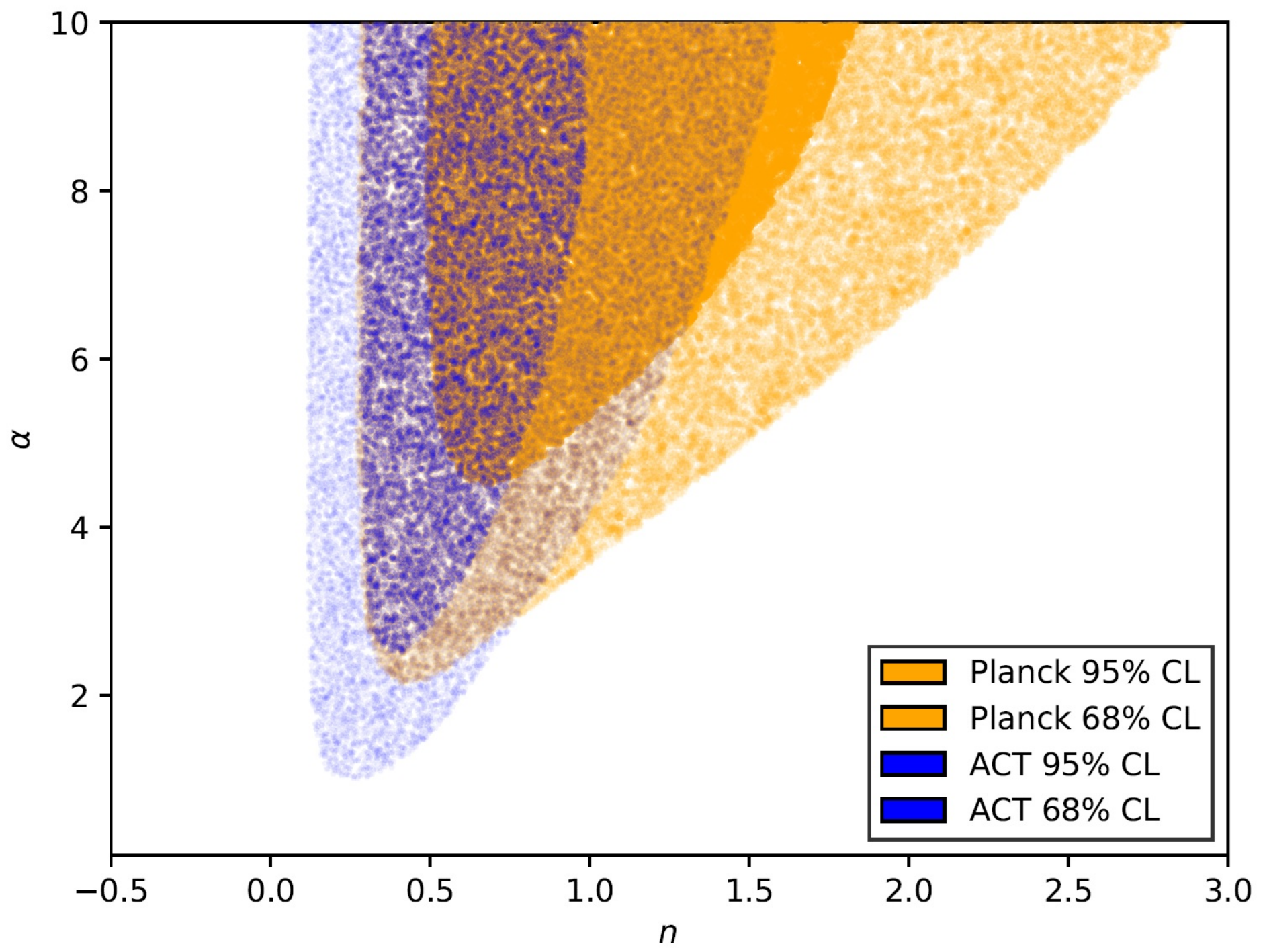}
\caption{Allowed region in the $(n,\alpha)$ plane consistent with observations. Orange contours: Planck 2018 ($95\%/68\%$ CL). Blue contours: ACT DR6 (2025) ($95\%/68\%$ CL).}
\label{pl_rns_param}
\end{figure}

	Figure~\ref{pl_rns_vs}(a) shows that for $n=0.5$, the predictions lie inside the ACT DR6 $68\%$ CL region and on the edge of the Planck contours. Increasing $n$ shifts $n_s$ to smaller values, moving the predictions outside the ACT-preferred range while remaining compatible with Planck. The complementary scan in Fig.~\ref{pl_rns_vs}(b) indicates that smaller $\alpha$ values are favored by ACT, whereas Planck allows somewhat larger $\alpha$. The overlap of the two datasets defines the viable parameter space shown in Fig.~\ref{pl_rns_param}. Overall, compatibility with ACT requires a tighter interval for $n$ than Planck, which permits a broader range.
In addition to the spectral observables, the amplitude of scalar perturbations $\mathcal{P}_s$ provides a normalization condition, allowing us to fix $\xi$:
\begin{align}\label{xi_powerlaw}
\xi^{2n} = \frac{1}{H_0^{4\alpha}} 
\left( \left[ \frac{\alpha (1+\lambda)}{2^\alpha M_p^2} \right]^{\tfrac{1}{2\alpha - 1}} 
\frac{\epsilon_1^\star}{n^{\tfrac{2\alpha}{2\alpha - 1}}} \right)^{2n(2\alpha - 1)} 
\left( \frac{1}{8 \pi^2 M_p^2 c_s \epsilon_1^\star \mathcal{P}_s}\right)^{2n(1-\alpha) - 2\alpha}.
\end{align}
Here, $M$ denotes the mass scale in the noncanonical kinetic term (see Eq.~\eqref{action}), while the inflationary energy scale is identified with $\rho^{1/4}$ at horizon crossing.
\begin{table}[h]
\centering
\footnotesize
\caption{Numerical results for the power-law case: scalar spectral index $n_s$, tensor-to-scalar ratio $r$, constant $M$, and inflationary energy scale (ES [$10^{-3}$]) for various $(\alpha,n)$ with $N=65$ and $\lambda=0.1$. All dimensional quantities are in Planck units.}
\setlength{\tabcolsep}{7pt} 
\renewcommand{\arraystretch}{1.2} 
\begin{tabular}{lcccccc}
	\hline\hline
	$n$ & $\alpha$ & $c_s$ & $n_s$ & $r [10^{-2}]$ & $M$ & ES[$10^{-3}$] \\
	\hline
	$0.5$ & $3.5$ & $0.3523$ & $0.9750$ & $2.725$ & $3.085 \times 10^{-11}$ & $2.407$ \\
	$0.5$ & $4.0$ & $0.3162$ & $0.9749$ & $2.465$ & $5.956 \times 10^{-11}$ & $2.347$ \\
	$0.5$ & $4.5$ & $0.2863$ & $0.9749$ & $2.244$ & $9.370 \times 10^{-11}$ & $2.293$ \\
	$0.5$ & $5.0$ & $0.2607$ & $0.9748$ & $2.052$ & $1.298 \times 10^{-10}$ & $2.242$ \\
	$0.5$ & $5.5$ & $0.2383$ & $0.9748$ & $1.883$ & $1.650 \times 10^{-10}$ & $2.194$ \\
	$0.8$ & $3.5$ & $0.3523$ & $0.9719$ & $3.607$ & $5.694 \times 10^{-9}$ & $2.581$ \\
	$0.8$ & $4.0$ & $0.3162$ & $0.9719$ & $3.248$ & $8.779 \times 10^{-9}$ & $2.515$ \\
	$0.8$ & $4.5$ & $0.2863$ & $0.9718$ & $2.948$ & $1.180 \times 10^{-8}$ & $2.455$ \\
	$0.8$ & $5.0$ & $0.2607$ & $0.9718$ & $2.690$ & $1.457 \times 10^{-8}$ & $2.399$ \\
	$0.8$ & $5.5$ & $0.2383$ & $0.9718$ & $2.462$ & $1.698 \times 10^{-8}$ & $2.346$ \\
	$1.0$ & $3.5$ & $0.3523$ & $0.9695$ & $4.302$ & $7.249 \times 10^{-8}$ & $2.698$ \\
	$1.0$ & $4.0$ & $0.3162$ & $0.9695$ & $3.862$ & $1.002 \times 10^{-7}$ & $2.626$ \\
	$1.0$ & $4.5$ & $0.2863$ & $0.9695$ & $3.497$ & $1.248 \times 10^{-7}$ & $2.561$ \\
	$1.0$ & $5.0$ & $0.2607$ & $0.9695$ & $3.184$ & $1.457 \times 10^{-7}$ & $2.502$ \\
	$1.0$ & $5.5$ & $0.2383$ & $0.9695$ & $2.910$ & $1.629 \times 10^{-7}$ & $2.447$ \\
	$1.2$ & $3.5$ & $0.3523$ & $0.9675$ & $4.866$ & $3.219 \times 10^{-7}$ & $2.782$ \\
	$1.2$ & $4.0$ & $0.3162$ & $0.9675$ & $4.356$ & $4.174 \times 10^{-7}$ & $2.706$ \\
	$1.2$ & $4.5$ & $0.2863$ & $0.9676$ & $3.936$ & $4.973 \times 10^{-7}$ & $2.638$ \\
	$1.2$ & $5.0$ & $0.2607$ & $0.9676$ & $3.579$ & $5.619 \times 10^{-7}$ & $2.576$ \\
	$1.2$ & $5.5$ & $0.2383$ & $0.9676$ & $3.267$ & $6.129 \times 10^{-7}$ & $2.518$ \\
	$1.5$ & $3.5$ & $0.3523$ & $0.9658$ & $5.331$ & $8.498 \times 10^{-7}$ & $2.846$ \\
	$1.5$ & $4.0$ & $0.3162$ & $0.9659$ & $4.763$ & $1.057 \times 10^{-6}$ & $2.767$ \\
	$1.5$ & $4.5$ & $0.2863$ & $0.9660$ & $4.296$ & $1.224 \times 10^{-6}$ & $2.697$ \\
	$1.5$ & $5.0$ & $0.2607$ & $0.9661$ & $3.901$ & $1.354 \times 10^{-6}$ & $2.632$ \\
	$1.5$ & $5.5$ & $0.2383$ & $0.9661$ & $3.557$ & $1.453 \times 10^{-6}$ & $2.572$ \\
	\hline\hline
\end{tabular}
\label{first_case_data}
\end{table}

\begin{figure}[h]
\centering
\begin{minipage}{0.48\textwidth}
	\centering
	\includegraphics[width=\linewidth]{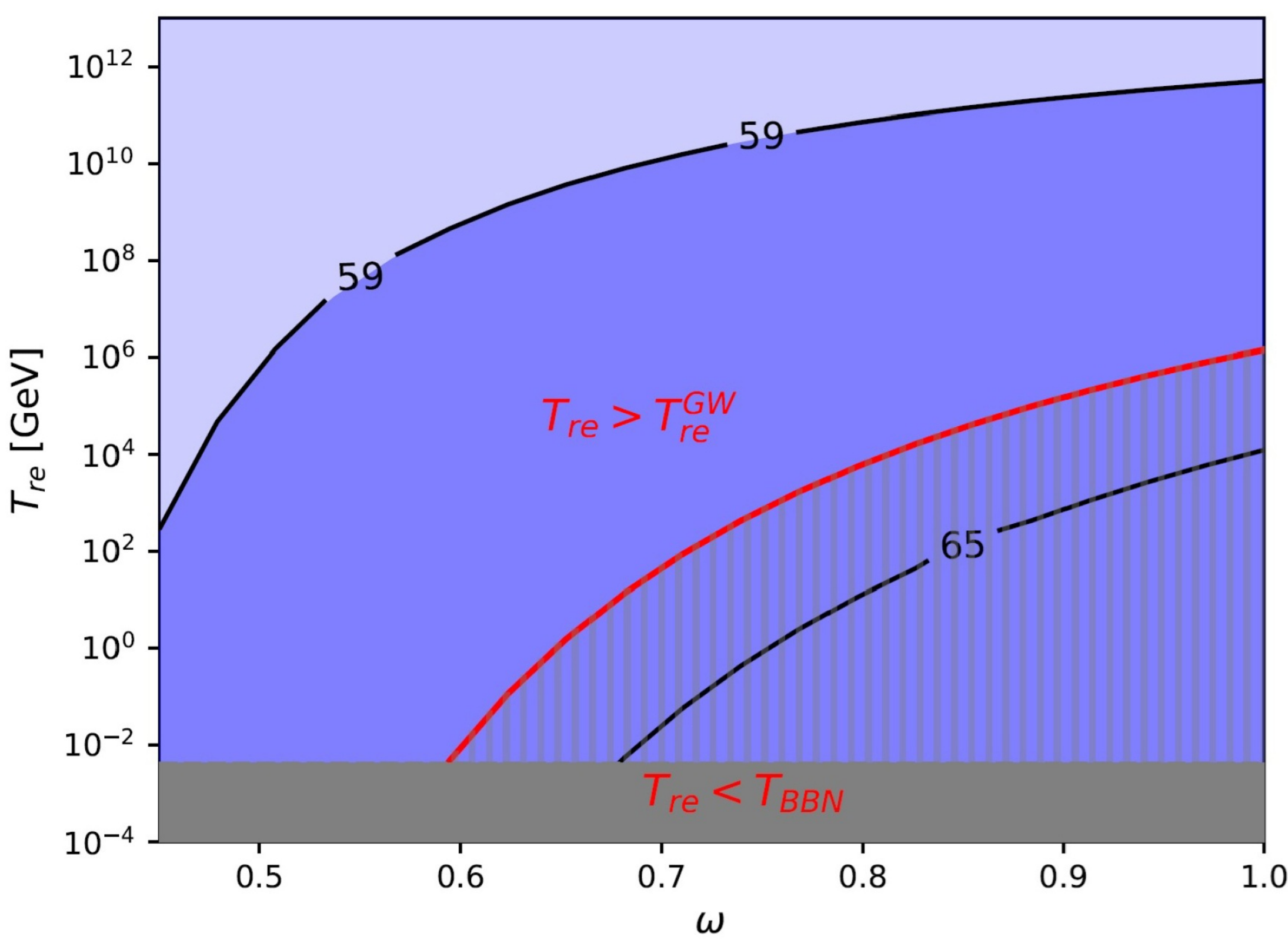}
	(a)\label{Tre_PL_1}
\end{minipage}
\hfill
\begin{minipage}{0.48\textwidth}
	\centering
	\includegraphics[width=\linewidth]{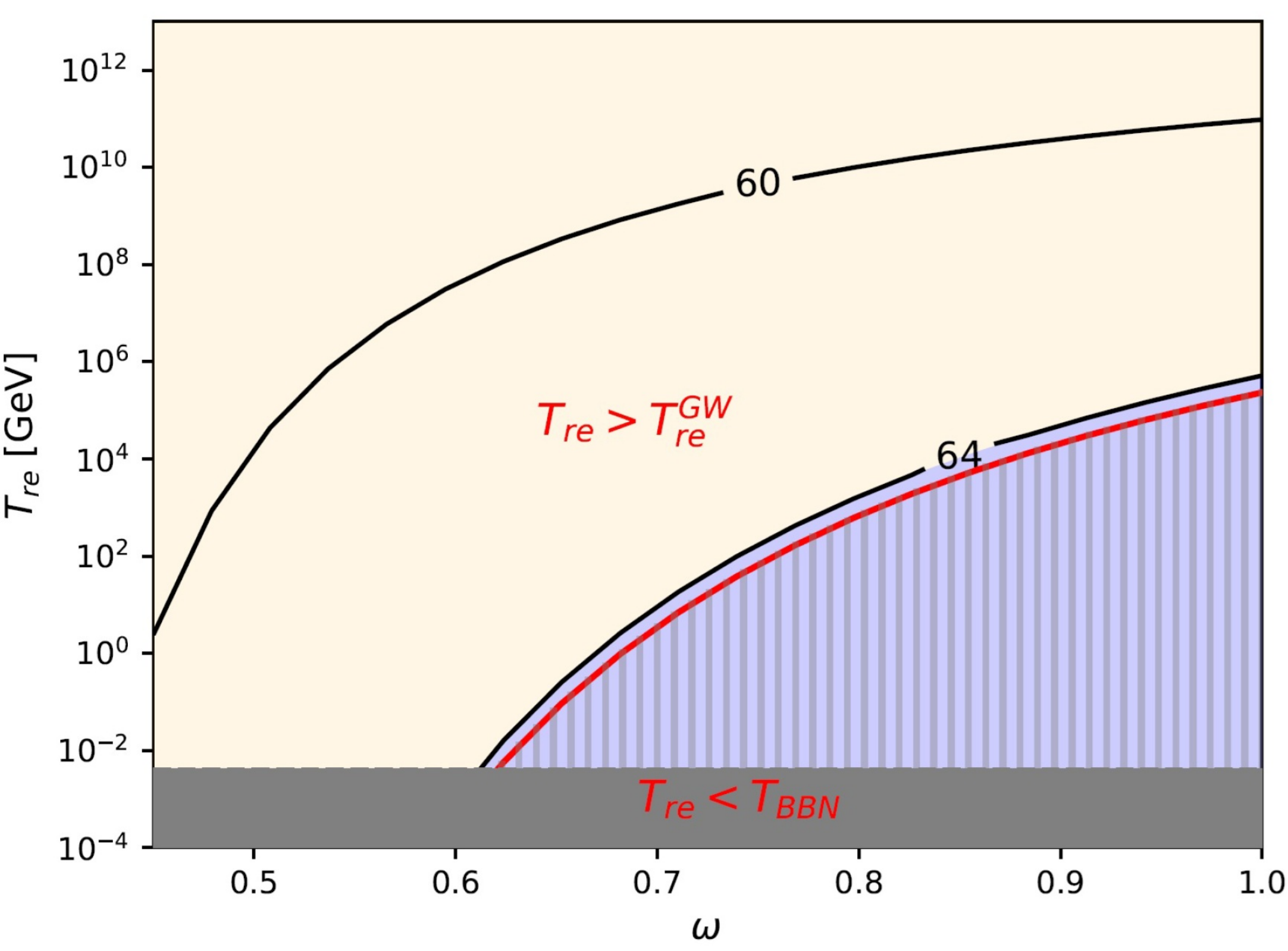}
	(b)\label{Tre_PL_2}
\end{minipage}
\caption{Reheating temperature $T_{\text{re}}$ versus $\omega_{\text{re}}$ for 
	(a) $(n,\alpha)=(0.5,3.5)$ and 
	(b) $(n,\alpha)=(1.2,6)$. Shaded bands: ACT DR6 (2025) $95\%/68\%$ CL in blue and Planck 2018 $95\%$ CL in orange. The red curve indicates the lower bound $T_{\text{re}}^{\rm GW}$ from the PGW constraint $\Delta N_{\rm eff}\leq0.17$ (effective for $\omega_{\text{re}}\gtrsim 0.59$); the region below the red curve is excluded.}
\label{Tre_PL}
\end{figure}

The reheating dynamics follow directly from Eq.~\eqref{Tre2}, showing that the reheating temperature $T_{\text{re}}$ is tightly correlated with the number of $e$-folds and the inflationary energy scales. The allowed window spans from the BBN limit $T_{BBN}\simeq4\,\mathrm{MeV}$ to the GUT scale $T_{GUT}\simeq10^{16}\,\mathrm{GeV}$. Importantly, tensor modes re-entering the horizon during reheating enhance the PGW background (more strongly for $\omega_{\text{re}}>1/3$), thereby contributing to $\Delta N_{\rm eff}$. The combined Planck+ACT bound, $\Delta N_{\rm eff}\leq0.17$ at $95\%$ CL, imposes an additional constraint on $T_{\text{re}}$, illustrated in Fig.~\ref{Tre_PL}.
As seen in Fig.~\ref{Tre_PL}(a), the case $(n,\alpha)=(0.5,3.5)$ remains consistent with ACT for $N\in[59,63]$, while larger $N$ violates the PGW reheating bound. For $(n,\alpha)=(1.2,6)$, agreement with Planck at $95\%$ CL is achievable, but ACT compatibility is lost because the required $N$ is excluded by $T_{\text{re}}^{\rm GW}$. Thus, reheating considerations further restrict the inflationary parameter space.

\begin{figure}[h]
\centering
\begin{minipage}{0.48\textwidth}
	\centering
	\includegraphics[width=\linewidth]{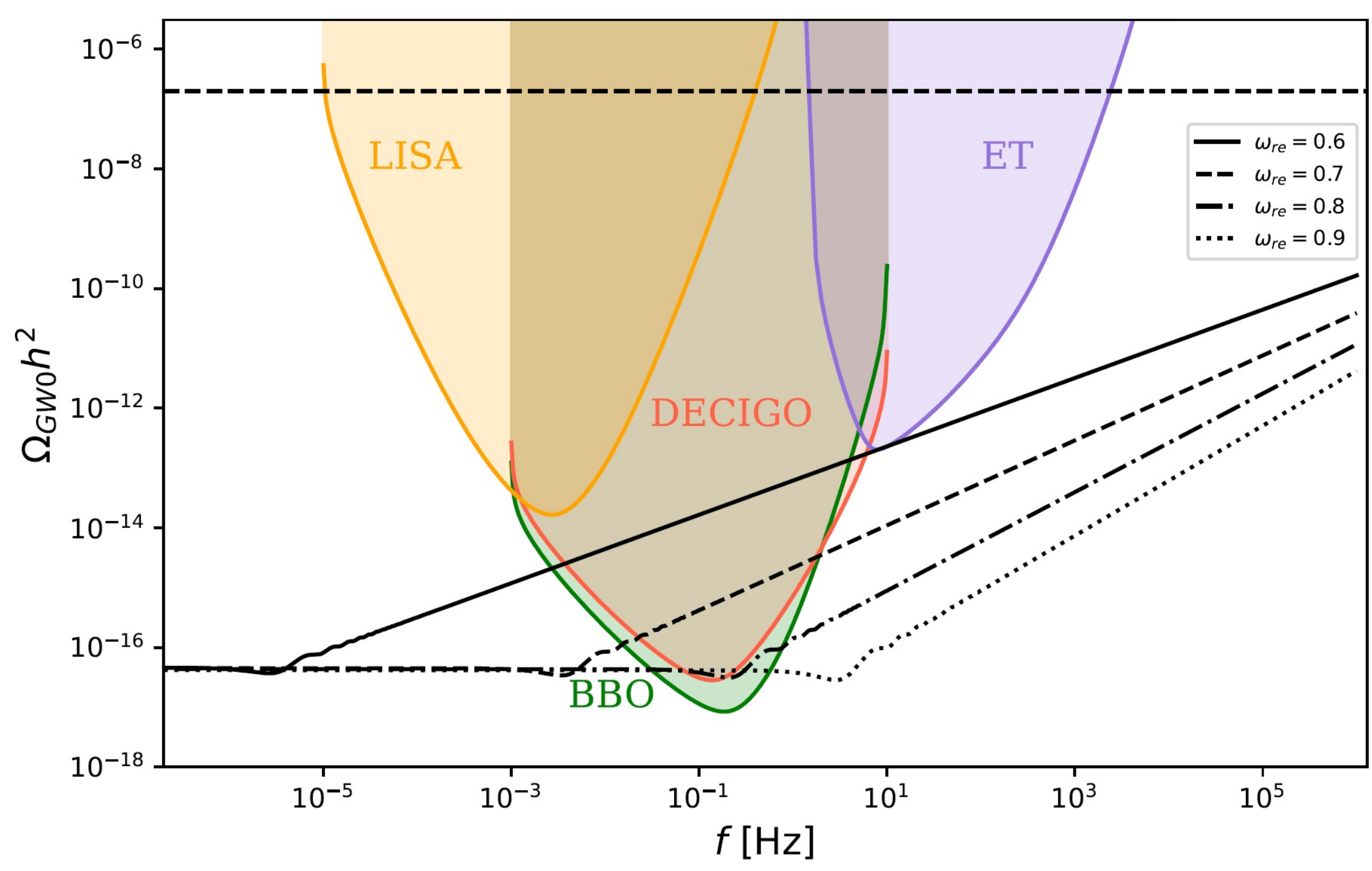}
	(a)\label{gw1}
\end{minipage}
\hfill
\begin{minipage}{0.48\textwidth}
	\centering
	\includegraphics[width=\linewidth]{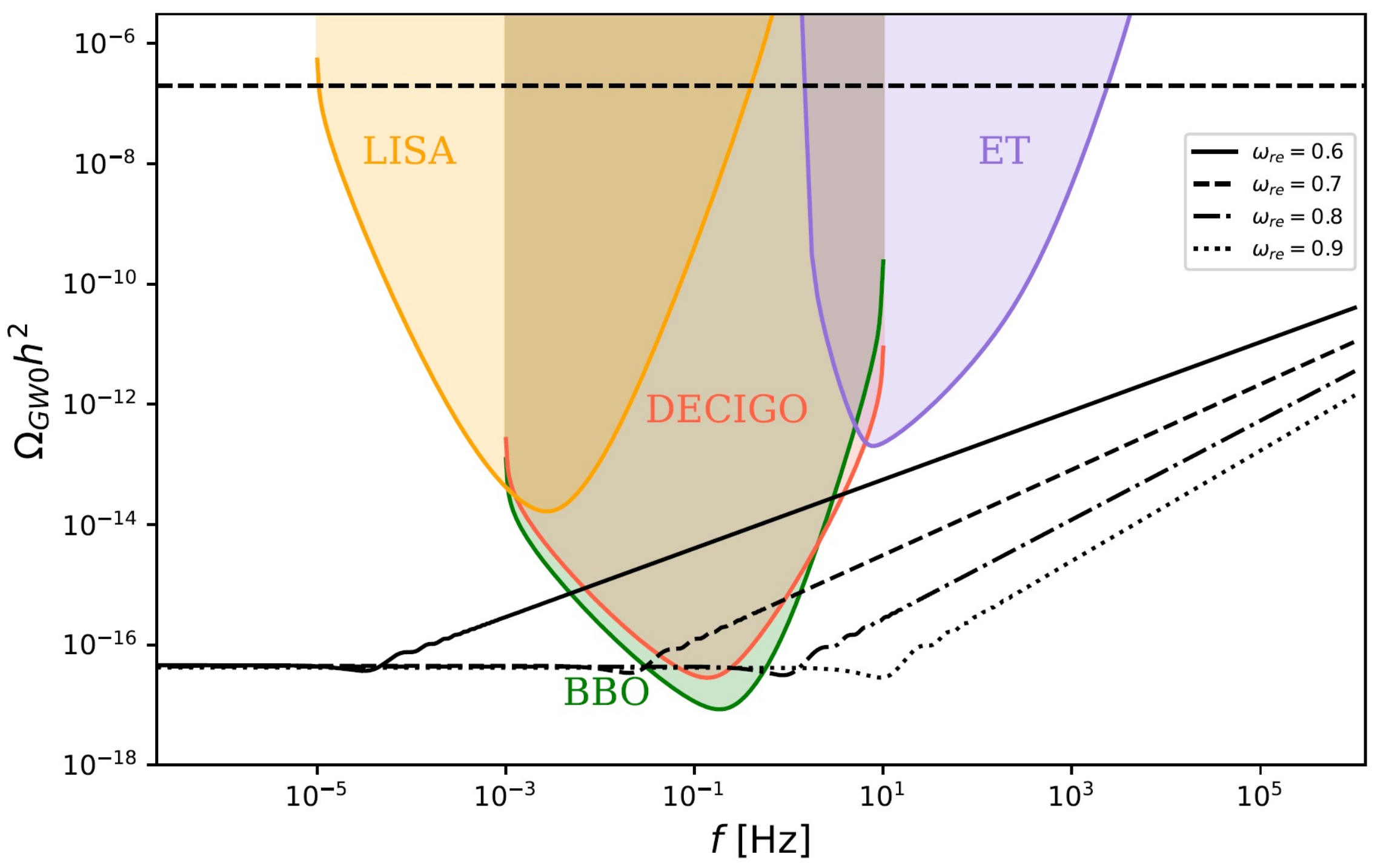}
	(b)\label{gw2}
\end{minipage}
\caption{Present-day PGW spectrum $\Omega_{\rm GW}(f)$ for several $\omega_{\text{re}}$ values and two inflationary durations: (a) $N_k=60$, (b) $N_k=65$.}
\label{gw}
\end{figure}

Finally, the present-day PGW spectrum is shown in Fig.~\ref{gw}. It is approximately scale-invariant at low frequencies and develops a blue tilt at higher frequencies; increasing $\omega_{\text{re}}$ shifts the enhancement to higher frequencies. Comparing $N_k=60$ and $N_k=65$ in Figs.~\ref{gw}(a) and \ref{gw}(b), a longer inflationary phase shifts the enhancement to higher frequencies, as more short-wavelength modes re-enter during reheating.

\subsection{Case II: Exponential inflation}

\begin{figure}[h]
\centering
\includegraphics[width=0.5\linewidth]{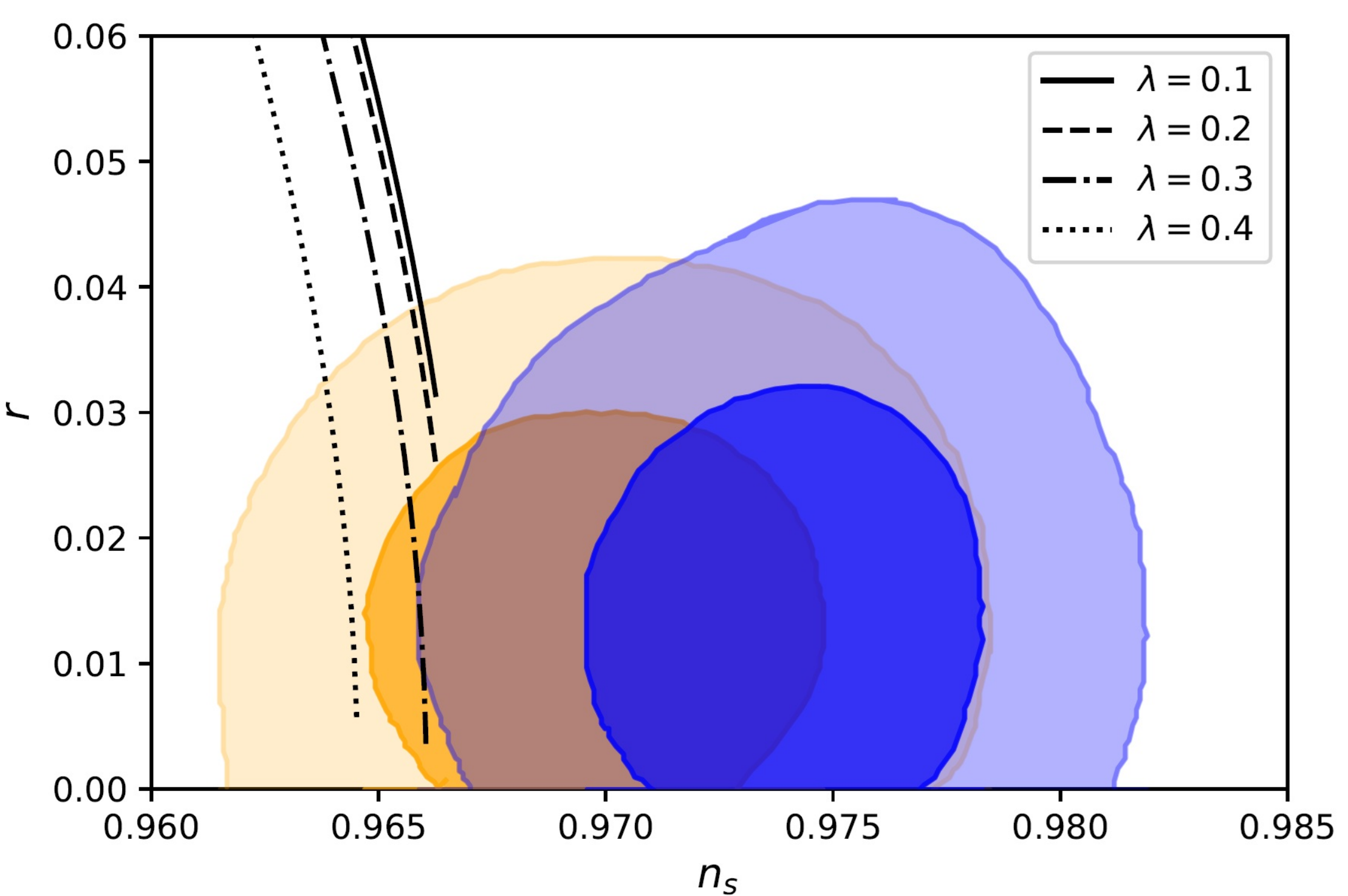}
\caption{Predictions of the exponential model in the $r$--$n_s$ plane. 
	Trajectories are shown as functions of $\alpha$ and for different values of $\lambda$, with $N=90$. Shaded regions denote the $68\%$ and $95\%$ CL bounds from Planck (orange) and ACT DR6 (blue).}
\label{exp_rns_vs_a}
\end{figure}

\begin{figure}[h]
\centering
\includegraphics[width=0.5\linewidth]{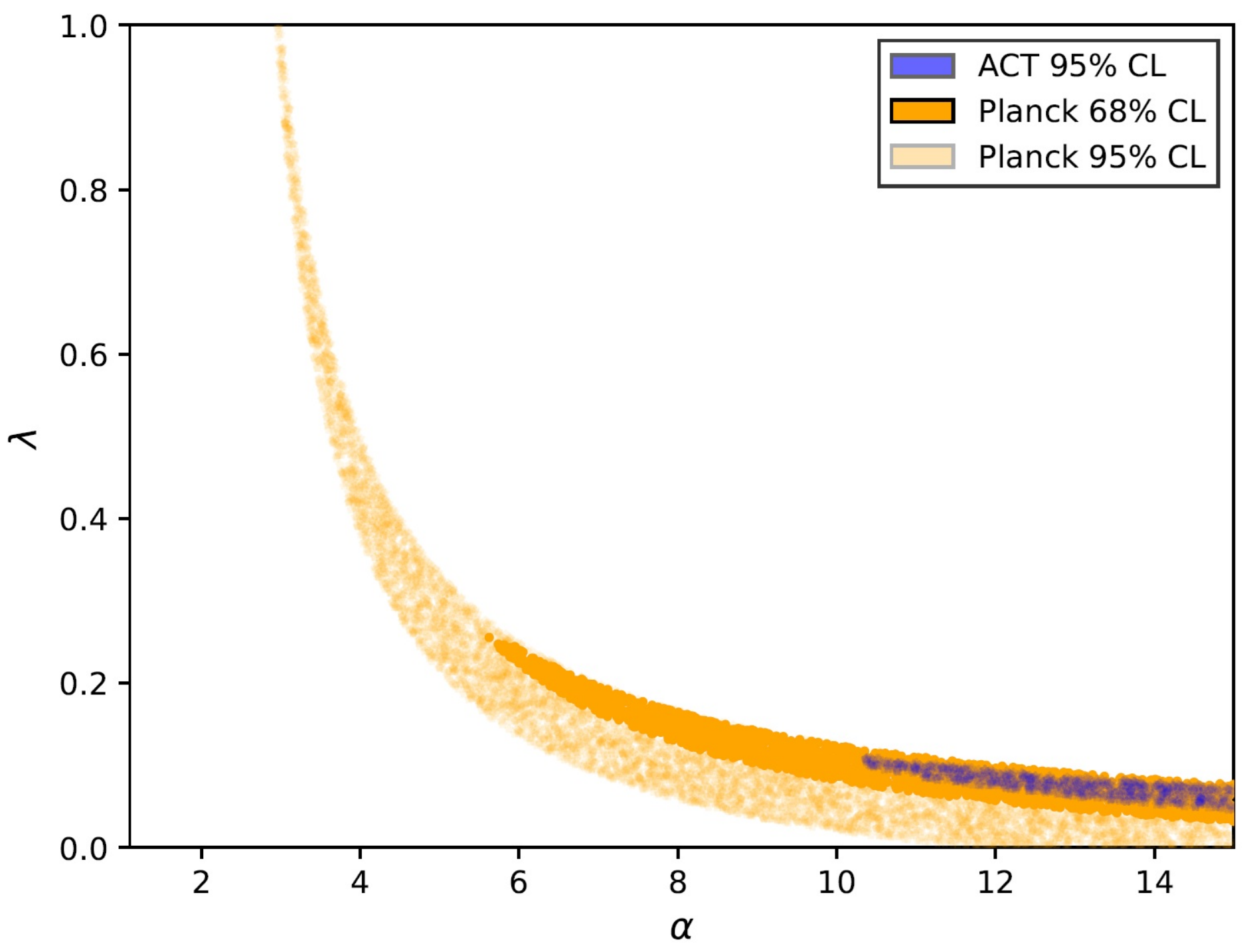}
\caption{Allowed regions in the parameter space $(\alpha,\lambda)$ for the exponential case with $N=90$. The light/dark orange regions indicate compatibility with Planck at $95\%$/$68\%$ CL, and the light blue region indicates compatibility with ACT DR6 at $95\%$ CL.}
\label{exp_rns_param}
\end{figure}

As a second example, we consider an exponential dependence of the Hubble parameter on the scalar field,
\begin{equation}
H(\phi) = H_0 e^{\beta \phi},
\end{equation}
where $H_0$ and $\beta$ are positive constants. Substituting this form into the Hamilton-Jacobi relations, the slow-roll parameters become
\begin{eqnarray}
\epsilon_1 &=& \left( \frac{2^\alpha M_p^2 \xi}{\alpha (1+\lambda)} \, \beta^{2\alpha} H_0^{2-2\alpha} \right)^{\tfrac{1}{2\alpha - 1}} 
\exp\!\left(\tfrac{2-2\alpha}{2\alpha - 1} \, \beta \phi \right), \label{srp1_exp} \\
\eta_H &=& \frac{1}{2\alpha - 1} 
\left( \frac{2^\alpha M_p^2 \xi}{\alpha (1+\lambda)} \, \beta^{2\alpha} H_0^{2-2\alpha} \right)^{\tfrac{1}{2\alpha - 1}} 
\exp\!\left(\tfrac{2-2\alpha}{2\alpha - 1} \, \beta \phi \right). \label{srp2_exp}
\end{eqnarray}
Inflation ends when $\epsilon_1=1$, which determines the field value at the end of inflation as
\begin{equation}\label{phi_end_exp}
\exp\!\left(\tfrac{2\alpha - 2}{2\alpha - 1} \, \beta \phi_e \right) 
= \left( \frac{2^\alpha M_p^2 \xi}{\alpha (1+\lambda)} \, \beta^{2\alpha} H_0^{2-2\alpha} \right)^{\tfrac{1}{2\alpha - 1}}.
\end{equation}
The field value at horizon crossing $\phi_\star$ is related to $\phi_e$ via the number of $e$-folds $N$. Integration of Eq.~\eqref{efolds} yields
\begin{equation}\label{phi_star_exp}
\exp\!\left(\tfrac{2\alpha - 2}{2\alpha - 1} \, \beta \phi_\star \right)
= \left( 1 + \tfrac{2\alpha - 2}{2\alpha - 1} N \right) 
\exp\!\left(\tfrac{2\alpha - 2}{2\alpha - 1} \, \beta \phi_e \right).
\end{equation}
Hence, the slow-roll parameters at horizon exit simplify to
\begin{eqnarray}
\epsilon_1^\star &=& \left( 1 + \tfrac{2\alpha - 2}{2\alpha - 1} N \right)^{-1}, \\
\eta_H^\star &=& \frac{1}{2\alpha -1} \, \epsilon_1^\star.
\end{eqnarray}
Using these, the scalar spectral index $n_s$ and the tensor-to-scalar ratio $r$ follow from Eqs.~\eqref{ns} and \eqref{r}. The resulting $r$--$n_s$ trajectories for varying $\alpha$ and several $\lambda$ are shown in Fig.~\ref{exp_rns_vs_a} for $N=90$. For $\lambda=0.3$, the curve can intersect the Planck $68\%$ CL region, whereas larger $\lambda$ shifts $n_s$ and prevents entry into that region (e.g., $\lambda=0.4$). None of the trajectories cross the ACT DR6 allowed region for the considered range of parameters.
A complementary scan in the $(\alpha,\lambda)$ plane is presented in Fig.~\ref{exp_rns_param} for $N=90$. The model can be made consistent with Planck, but only a very narrow slice reaches the ACT $95\%$ CL boundary; the $68\%$ CL region of ACT is not attained for the displayed range. As $\alpha$ decreases, achieving agreement requires larger $\lambda$, while the viable interval of $\lambda$ becomes increasingly narrow.
\begin{table}[h]
\centering
\footnotesize
\caption{Numerical results for the exponential case: scalar spectral index $n_s$, tensor-to-scalar ratio $r$, coupling $\xi$, and inflationary energy scale (ES) for different $(\alpha,\lambda)$ with $N=90$. All dimensional quantities are in Planck units.}
\setlength{\tabcolsep}{7pt} 
\renewcommand{\arraystretch}{1.2} 
\begin{tabular}{lcccccc}
	\hline\hline
	$\alpha$ & $\lambda$ & $c_s$ & $n_s$ & $r [10^{-2}]$ & $M [10^{-5}]$ & ES[$10^{-3}$] \\
	\hline
	$10.0$ & $0.0$ & $0.2184$ & $0.9658$ & $4.052$ & $7.557$ & $2.658$ \\
	$10.0$ & $0.0$ & $0.1957$ & $0.9658$ & $3.630$ & $7.357$ & $2.586$ \\
	$10.0$ & $0.1$ & $0.1715$ & $0.9658$ & $3.181$ & $7.122$ & $2.502$ \\
	$10.0$ & $0.1$ & $0.1302$ & $0.9658$ & $2.415$ & $6.653$ & $2.335$ \\
	$11.0$ & $0.0$ & $0.2066$ & $0.9660$ & $3.813$ & $7.500$ & $2.618$ \\
	$11.0$ & $0.0$ & $0.1824$ & $0.9660$ & $3.366$ & $7.273$ & $2.537$ \\
	$11.0$ & $0.1$ & $0.1562$ & $0.9660$ & $2.882$ & $6.999$ & $2.441$ \\
	$11.0$ & $0.1$ & $0.1092$ & $0.9660$ & $2.015$ & $6.405$ & $2.232$ \\
	$12.0$ & $0.0$ & $0.1964$ & $0.9661$ & $3.608$ & $7.441$ & $2.582$ \\
	$12.0$ & $0.0$ & $0.1707$ & $0.9661$ & $3.136$ & $7.188$ & $2.493$ \\
	$12.0$ & $0.1$ & $0.1423$ & $0.9661$ & $2.614 $ & $6.871$ & $2.382$ \\
	$12.0$ & $0.1$ & $0.0882$ & $0.9661$ & $1.620$ & $6.101$ & $2.113$ \\
	$13.0$ & $0.0$ & $0.1873$ & $0.9661$ & $3.429$ & $7.383$ & $2.549$ \\
	$13.0$ & $0.0$ & $0.1602$ & $0.9661$ & $2.933$ & $7.102$ & $2.451$ \\
	$13.0$ & $0.1$ & $0.1295$ & $0.9661$ & $2.370$ & $6.737$ & $2.324$ \\
	$13.0$ & $0.1$ & $0.0655$ & $0.9661$ & $1.199$ & $5.685$ & $1.960$ \\
	\hline\hline
\end{tabular}
\label{table_exp}
\end{table}
The normalization condition from the scalar perturbation amplitude $\mathcal{P}_s$ fixes the coupling $\xi$,
\begin{equation}
\xi = \frac{\alpha (1+\lambda)}{2^\alpha M_p^2} \,
\frac{(\epsilon_1^\star)^{2\alpha - 1}}{\big( 8 \pi^2 M_p^2 c_s \epsilon_1^\star \mathcal{P}_s \big)^{1-\alpha} \, \beta^{2\alpha}}.
\end{equation}
From this, the mass scale $M=\xi^{1/[4(\alpha-1)]}$ can be extracted and is typically of order $\mathcal{O}(10^{-5})$ in Planck units ($M_p=1$), consistent with $M \ll M_p$.

\begin{figure}
\centering
\includegraphics[width = 0.5\linewidth]{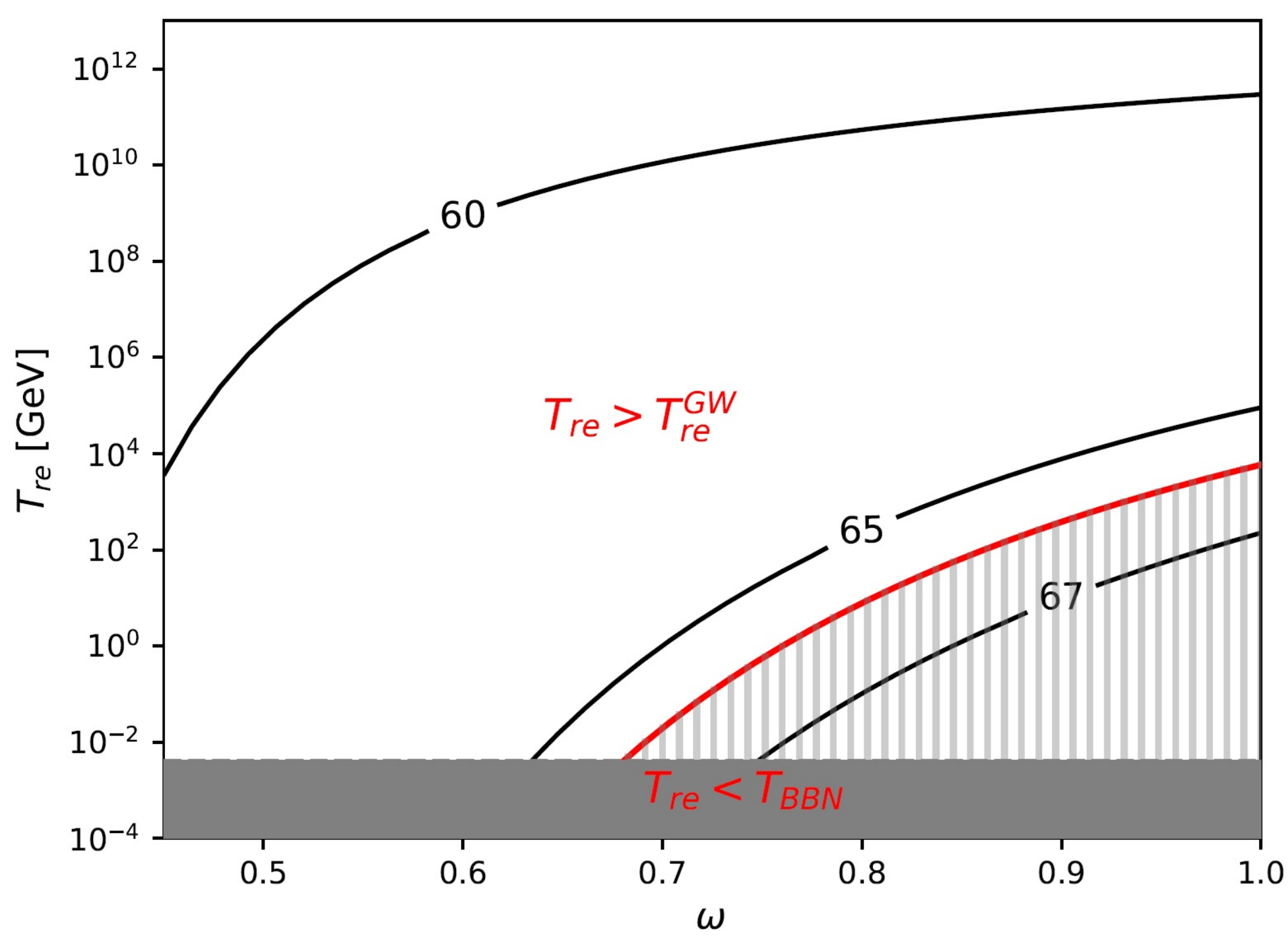}
\caption{\label{Tre_exp}
	Reheating temperature versus the effective reheating equation of state $\omega_{\text{re}}$ for $\alpha = 12$ and $\lambda = 0.08$. The red line denotes $T_{\text{re}}^{GW}$; the shaded region below violates $T_{\text{re}} > T_{\text{re}}^{GW}$. Satisfying the bound requires $N<66$, but such values yield $(n_s,r)$ inconsistent with data.}
	\end{figure}

To further assess viability, we examine the reheating temperature $T_{\text{re}}$ versus the effective reheating equation-of-state parameter $\omega_{\text{re}}$ in Fig.~\ref{Tre_exp}, for a representative choice $\alpha=12$ and $\lambda=0.08$. Increasing $N$ lowers $T_{\text{re}}$; the red line indicates the lower bound $T_{\text{re}}^{GW}$ from the PGW constraint $\Delta N_{\rm eff}\leq 0.17$. For $N \gtrsim 66$, the temperature bound is violated, and on the other hand, for $N \lesssim 66$ moves $(n_s,r)$ outside the observationally allowed ranges. Hence, within this setup, the exponential case is disfavored.
	\clearpage
	\section{Swampland Criteria and TCC}\label{SC-TCC} 
	
	String theory, as one of the most promising candidates for a consistent theory of quantum gravity, predicts a vast ``landscape'' of effective field theories (EFTs) that can be consistently embedded in UV-complete frameworks. In contrast, some EFTs, though apparently consistent at low energies, fail to admit such embeddings and are believed to lie in the "swampland''. To distinguish between these cases, a set of conjectured consistency conditions, the swampland criteria, have been proposed. The two most relevant ones for cosmology are:
	\begin{itemize}
\item \textbf{Distance Conjecture}: The field excursion during the EFT validity is bounded by
\begin{equation}\label{c1}
	\Delta \phi \leq c_1,
\end{equation}
where $c_1=\mathcal{O}(1)$ in reduced Planck units \cite{PhysRevD.98.086004,Garg:2018reu,Ooguri:2018Refined,Palti:2019Review}.

\item \textbf{de Sitter Conjecture}: The scalar potential must satisfy a lower bound on its slope \cite{Garg:2018reu,Ooguri:2018Refined,Kehagias:2018uem},
\begin{equation}\label{c2}
	\frac{|V_{,\phi}|}{V} \geq c_2,
\end{equation}
or, in its refined form,
\begin{equation}\label{c2_refined}
	\frac{|V_{,\phi}|}{V} \geq c_2, \qquad
	\frac{V_{,\phi\phi}}{V} \geq -c_2',
\end{equation}
with $c_2,c_2'>0$. The precise values depend on compactification details, but the key requirement is their positivity, typically $\mathcal{O}(0.1$--$1)$.
\end{itemize}
Inflation occurs well below the Planck scale, making its description as an EFT natural. It is therefore crucial to check whether inflationary scenarios are consistent with the above criteria. Using the observationally constrained parameters from the previous section, we compute the field values at horizon crossing and at the end of inflation, and evaluate both the excursion $\Delta \phi = |\phi_\star-\phi_e|$ and the slope $V'/V$. The results are summarized in Tables~\ref{table_sc_A} and \ref{table_sc_B} for the power law and the exponential case, respectively.
\begin{table}[h]
\centering
\footnotesize
\caption{Field excursion $\Delta \phi$ and potential gradient $V'/V$ for the power-law case with various values of $\alpha$ and $n$, taking $\lambda = 0.1$ and $N=65$. For the acceptable parameter values, both swampland criteria are perfectly satisfied.}
\setlength{\tabcolsep}{7pt}
\renewcommand{\arraystretch}{1.2}
\begin{tabular}{lccc}
	\hline\hline
	$\alpha$ & $n$ & $\Delta \phi$ & $V'/V$ \\
	\hline
	$0.5$ & $3.5$ & $2.699 \times 10^{-10}$ & $3.593 \times 10^{9}$ \\
	$0.5$ & $4.0$ & $2.443 \times 10^{-10}$ & $3.973 \times 10^{9}$ \\
	$0.5$ & $4.5$ & $2.226 \times 10^{-10}$ & $4.364 \times 10^{9}$ \\
	$0.5$ & $5.0$ & $2.037 \times 10^{-10}$ & $4.772 \times 10^{9}$ \\
	$0.5$ & $5.5$ & $1.869 \times 10^{-10}$ & $5.203 \times 10^{9}$ \\
	$0.8$ & $3.5$ & $4.847 \times 10^{-7}$ & $2.924 \times 10^{6}$ \\
	$0.8$ & $4.0$ & $4.522 \times 10^{-7}$ & $3.135 \times 10^{6}$ \\
	$0.8$ & $4.5$ & $4.241 \times 10^{-7}$ & $3.345 \times 10^{6}$ \\
	$0.8$ & $5.0$ & $3.990 \times 10^{-7}$ & $3.556 \times 10^{6}$ \\
	$0.8$ & $5.5$ & $3.763 \times 10^{-7}$ & $3.772 \times 10^{6}$ \\
	$1.0$ & $3.5$ & $1.917 \times 10^{-5}$ & $9.563 \times 10^{4}$ \\
	$1.0$ & $4.0$ & $1.817 \times 10^{-5}$ & $1.009 \times 10^{5}$ \\
	$1.0$ & $4.5$ & $1.729 \times 10^{-5}$ & $1.061 \times 10^{5}$ \\
	$1.0$ & $5.0$ & $1.649 \times 10^{-5}$ & $1.112 \times 10^{5}$ \\
	$1.0$ & $5.5$ & $1.577 \times 10^{-5}$ & $1.163 \times 10^{5}$ \\
	$1.2$ & $3.5$ & $1.681 \times 10^{-4}$ & $1.320 \times 10^{4}$ \\
	$1.2$ & $4.0$ & $1.608 \times 10^{-4}$ & $1.380 \times 10^{4}$ \\
	$1.2$ & $4.5$ & $1.543 \times 10^{-4}$ & $1.437 \times 10^{4}$ \\
	$1.2$ & $5.0$ & $1.485 \times 10^{-4}$ & $1.493 \times 10^{4}$ \\
	$1.2$ & $5.5$ & $1.431 \times 10^{-4}$ & $1.549 \times 10^{4}$ \\
	$1.5$ & $3.5$ & $6.989 \times 10^{-4}$ & $3.685 \times 10^{3}$ \\
	$1.5$ & $4.0$ & $6.723 \times 10^{-4}$ & $3.827 \times 10^{3}$ \\
	$1.5$ & $4.5$ & $6.490 \times 10^{-4}$ & $3.961 \times 10^{3}$ \\
	$1.5$ & $5.0$ & $6.280 \times 10^{-4}$ & $4.090 \times 10^{3}$ \\
	$1.5$ & $5.5$ & $6.086 \times 10^{-4}$ & $4.218 \times 10^{3}$ \\
	\hline\hline
\end{tabular}
\label{table_sc_A}
\end{table}

\begin{table}[h]
\centering
\footnotesize
\caption{Field excursion $\Delta \phi$ and potential gradient $V'/V$ for the exponential case with different values of $\alpha$ and $\lambda$, taking $N=90$. Both swampland criteria are perfectly satisfied.}
\setlength{\tabcolsep}{7pt}
\renewcommand{\arraystretch}{1.2}
\begin{tabular}{lccc}
	\hline\hline
	$\alpha$ & $\lambda$ & $\Delta \phi$ & $V'/V$ \\
	\hline
	$10.0$ & $0.01$ & $0.9410$ & $10.0352$ \\
	$10.0$ & $0.03$ & $0.9410$ & $10.0355$ \\
	$10.0$ & $0.05$ & $0.9410$ & $10.0358$ \\
	$10.0$ & $0.08$ & $0.9410$ & $10.0363$ \\
	$11.0$ & $0.01$ & $0.9371$ & $10.0354$ \\
	$11.0$ & $0.03$ & $0.9371$ & $10.0357$ \\
	$11.0$ & $0.05$ & $0.9371$ & $10.0360$ \\
	$11.0$ & $0.08$ & $0.9371$ & $10.0365$ \\
	$12.0$ & $0.01$ & $0.9340$ & $10.0355$ \\
	$12.0$ & $0.03$ & $0.9340$ & $10.0358$ \\
	$12.0$ & $0.05$ & $0.9340$ & $10.0362$ \\
	$12.0$ & $0.08$ & $0.9340$ & $10.0366$ \\
	$13.0$ & $0.01$ & $0.9314$ & $10.0356$ \\
	$13.0$ & $0.03$ & $0.9314$ & $10.0360$ \\
	$13.0$ & $0.05$ & $0.9314$ & $10.0363$ \\
	$13.0$ & $0.08$ & $0.9314$ & $10.0367$ \\
	\hline\hline
\end{tabular}
\label{table_sc_B}
\end{table}
The results show that both conjectures are comfortably satisfied in the parameter ranges of interest.  
In the power-law case (Table~\ref{table_sc_A}), the field excursion is extremely small, $\Delta \phi \ll 1$, while the potential slope is very steep, $V'/V \gg 1$. This makes the Distance Conjecture trivially satisfied and the de Sitter bound easily fulfilled.  In the exponential case (Table~\ref{table_sc_B}), the excursion is still sub-Planckian, $\Delta \phi \simeq 0.93$, and the slope remains moderate but positive, $V'/V \sim 10$, consistent with the refined de Sitter conjecture. Importantly, these values are stable against changes in $\alpha$ and $\lambda$, highlighting the robustness of the results.

An additional condition is provided by the TCC \cite{Bedroya:2019TCC,Bedroya:2020TCCInflation}, which forbids sub-Planckian modes from ever crossing the Hubble horizon. It requires
\begin{equation}\label{tcc}
\frac{l_p}{a_i} < \frac{H_f^{-1}}{a_f},
\end{equation}
where $l_p$ is the Planck length, $H_f$ the Hubble scale at the end of inflation, and $a_i$, $a_f$ the scale factors at the beginning and end of inflation. In our models, typical values are $H_f^{-1} \sim \mathcal{O}(10^6)$ (Planck units) with $N \sim 65$ (power-law) or $N \sim 90$ (exponential). Since $e^N \gg H_f^{-1}$, inequality \eqref{tcc} is violated. Therefore, while both scenarios comfortably satisfy the swampland criteria, they remain in strong tension with the TCC, highlighting a fundamental challenge for embedding such inflationary models into a fully consistent UV-complete theory of quantum gravity. Importantly, this tension should not be interpreted as a complete exclusion of the models, but rather as evidence of the broader difficulty in reconciling inflationary cosmology with the foundational principles of effective field theory under quantum gravity constraints. This perspective leaves room for possible resolutions, such as modified reheating dynamics, alternative EFT embeddings, or string-theoretic corrections.

\clearpage

\section{Conclusions and Outlook}\label{conclusion}

In this work, we developed the Hamilton-Jacobi framework for noncanonical inflation in $f(R,T)$ gravity and studied two representative cases of the Hubble parameter: a power-law form and an exponential form. For both cases, we derived the slow-roll parameters, computed the scalar spectral index $n_s$ and the tensor-to-scalar ratio $r$ at horizon crossing, and compared the predictions with the latest observational bounds from Planck and ACT DR6. After inflation, the universe undergoes reheating, during which particles are produced, and it undergoes a smooth transition to the radiation-dominated phase. The reheating dynamics are characterized by the reheating temperature $T_{\text{re}}$, the number of reheating e-folds $N_{\text{re}}$, and the effective equation-of-state parameter $\omega_{\text{re}}$. We showed that reheating is closely linked to the inflationary phase. The reheating temperature depends on both the number of inflationary e-folds and the energy scales at horizon crossing and at the end of inflation. Moreover, the overproduction of primordial gravitational waves (PGWs), generated as tensor perturbations during inflation, imposes an additional lower bound on the reheating temperature through its effect on the effective number of relativistic species, $\Delta N_{\text{eff}}$. The BBN and CMB observational bound $\Delta N_{\text{eff}} < 0.17$ provides a lower bound for the reheating temperature, which is mostly efficient for the case of stiff reheating equation of state. Then, we calculated the resulting gravitational wave energy spectrum using the model parameters obtained within this framework.

Our analysis indicates that the power-law model is in good agreement with the data over a broad range of parameters. In particular, the benchmark point $(n,\alpha) = (0.5, 3.5)$ with $N \simeq 65$ yields $n_s \simeq 0.975$ and $r \simeq 0.0027$, which lies well within the joint $68\%$ CL region of ACT. It was determined that increasing the parameter $n$ leads to lower values of the scalar spectral index and higher values of the tensor-to-scalar ratio, whereas increasing $\alpha$ suppresses the tensor-to-scalar ratio. Using the allowed parameter space, we then studied the reheating temperature. Our results indicate that in order to satisfy the bound $T_{\text{re}} > T_{\text{re}}^{GW}$, the total number of inflationary e-folds should not exceed $64$–$65$ (with the precise value depending on the model). Thus, reheating can impose an additional constraint on the parameter space. We also examined the resulting gravitational wave energy spectrum. The amplitude of the spectrum increases with frequency, and this enhancement occurs at higher frequencies for larger values of $\omega_{\text{re}}$. For smaller  $\omega_{\text{re}}$, the spectrum can enter the observable range of future detectors such as BBO and DECIGO. Furthermore, a smaller number of e-folds extends the range of frequencies entering the observable window, thereby improving the prospects for detection.

Exploring the parametric space of the free parameters for the exponential function of the Hubble reveals that the model can roughly agree with the ACT data; however, it falls within the observable range of Planck. However, even this agreement occurs for high values of the number of e-folds, i.e., $N \geq 90$. Additionally, it is shown that for this range of number of e-folds, the reheating temperature could not satisfy the constraint $T_{\text{re}} > T_{\text{re}}^{GW}$. Due to this result, it is concluded that this case can not provide a reliable description of inflation.

We also examined the Swampland criteria, finding that both models satisfy the Distance and de Sitter conjectures across the observationally favored parameter space. However, both scenarios remain in strong tension with the TCC, highlighting a fundamental challenge for embedding such inflationary models into a fully consistent UV-complete theory of quantum gravity. This tension should not be interpreted as a full exclusion of the models, but rather as an indication of the broader difficulty in reconciling inflationary cosmology with foundational principles of effective field theory.

\appendix	
\clearpage


\end{document}